\documentclass[twocolumn]{aastex631}

\newcommand{\angstrom}{\mbox{\normalfont\AA}}

\usepackage{soul}

\usepackage{CJK}
\usepackage{amsmath}

\received{May 15, 2025}

\submitjournal{ApJ}

\begin{document}

\title{The $z=7.08$ Quasar ULAS J1120+0641 May Never Reach a ``Normal'' Black Hole to Stellar Mass Ratio}

\correspondingauthor{Meredith Stone}
\email{meredithstone@arizona.edu}

\author[0000-0002-9720-3255]{Meredith A. Stone}
\affiliation{Steward Observatory, University of Arizona, 933 North Cherry Avenue, Tucson, AZ 85721, USA}

\author[0000-0003-2303-6519]{George H. Rieke}
\affiliation{Steward Observatory, University of Arizona,
933 North Cherry Avenue, Tucson, AZ 85721, USA}

\author[0000-0002-6221-1829]{Jianwei Lyu (\begin{CJK}{UTF8}{gbsn}吕建伟\end{CJK})}
\affiliation{Steward Observatory, University of Arizona,
933 North Cherry Avenue, Tucson, AZ 85721, USA}

\author[0000-0001-5097-6755]{Michael K. Florian}
\affiliation{Steward Observatory, University of Arizona,
933 North Cherry Avenue, Tucson, AZ 85721, USA}

\author[0000-0003-4565-8239]{Kevin N. Hainline}
\affiliation{Steward Observatory, University of Arizona,
933 North Cherry Avenue, Tucson, AZ 85721, USA}

\author[0000-0001-6561-9443]{Yang Sun}
\affiliation{Steward Observatory, University of Arizona,
933 North Cherry Avenue, Tucson, AZ 85721, USA}

\author[0000-0003-3307-7525]{Yongda Zhu}
\affiliation{Steward Observatory, University of Arizona,
933 North Cherry Avenue, Tucson, AZ 85721, USA}

\begin{abstract}

JWST observations of quasars in the Epoch of Reionization have revealed that many lie in host galaxies that are severely undermassive relative to the supermassive black holes. It is unclear how these systems will evolve to the tight local relation between stellar mass and black hole mass. We search for companions around the $z=7.08$ quasar ULAS J1120+0641 using JWST/NIRCam narrow, medium, and wide-band photometry to identify  [O~III] emitters at the quasar redshift, and explore the potential for growth of the host galaxy through future mergers. We find 22 sources near the quasar's redshift across our two 4.4 arcmin$^2$ fields, indicating that environment of ULAS J1120+0641 is strongly overdense in $z\sim7.1$ galaxies relative to the field. We estimate the potential future mass budget of the quasar host galaxy by summing the current stellar and gas masses of the quasar host and surrounding galaxies, correcting for incompleteness and selection effects. With no further black hole growth, ULAS J1120+0641 is unlikely to reach a $M_{\mathrm{BH}}/M_*$ ratio less than $\sim2.5\%$ at $z=0$, still much higher than typical for local galaxies. However, such  systems---a quiescent black hole in a low-luminosity galaxy---may have escaped detection locally if they are sufficiently distant.

\end{abstract}

\section{Introduction} \label{sec:intro}

Bright quasars---the sites of massive, fast-growing supermassive black holes (SMBHs)---provide a fascinating laboratory to study the coevolution of SMBHs and their host galaxies, the physics of gas transport from the largest to smallest scales, and the effects of SMBH feedback on galaxies. The extreme luminosities of quasars have allowed their redshift frontier to expand swiftly from the local Universe \citep{Schmidt1963} back to the Epoch of Reionization at $z\gtrsim5$ \citep{Bogdan2024, Fan2001, Mortlock2011, Fan2023}. However, the study of quasar hosts (outshined by the bright quasar point source) and the much fainter galaxies in their environments is significantly more difficult, especially at high redshift. Both areas of research have expanded significantly in recent years with JWST.

Quasars must accrete to produce their luminosity, indicating they are likely to live in gas-rich environments (at least, in the black hole's immediate neighborhood). On larger scales, though, recent JWST results have indicated that some luminous quasars at $z\gtrsim6$ live in undermassive host galaxies compared to the local Universe \citep{Onoue2024, Stone2024, Yue2024}, though with some exceptions \citep{Ding2023}. Unfortunately, the difficulty of measuring the stellar mass of a bright quasar's host galaxy, especially at high redshift, means only a small number of high-redshift quasars currently have robust stellar masses. With this small sample, we cannot yet determine whether the behavior of these quasars reflects a genuine change in the relationship between black hole mass and stellar mass ($M_{\mathrm{BH}}/M_*$) at early times, or is simply a consequence of selection biases \citep{YSun2025}.

However, the fact remains that such extreme galaxies, with $M_{\mathrm{BH}}/M_* \gg 1\%$, are not observed in the local Universe aside from a few niche cases where the host galaxy has lost significant mass to tidal stripping. If these quasar host galaxies are to grow enough to land on the observed local relationship between SMBH mass and galaxy mass by $z=0$, they must be able to form stars and merge with other galaxies in their environment. This fact, combined with predictions of the $\Lambda$CDM paradigm and results of cosmological simulations, means when we observe quasars at the end of the Epoch of Reionization, we expect to see them living in dense environments, surrounded by other galaxies \citep{Overzier2009a, Inayoshi2020, Volonteri2021}. 

For decades, observations have targeted high-redshift quasars in hopes of detecting galaxy overdensities surrounding them. At high redshift ($z\gtrsim5$), HST and ground-based optical/NIR telescopes can only access the rest-frame ultraviolet, and must rely on the Lyman break technique \citep{Adams2015, Husband2013, Morselli2014, Ota2018, Pudoka2024, Stiavelli2005, Zheng2006} or selection of Lyman-alpha emitters \citep{Chanchaiworawit2019, Harikane2019, Hennawi2015, Kashikawa2007, Overzier2009b}, with relatively few bands longward of the Lyman break, to identify candidate galaxies. Although galaxies were detected around many quasars, it remained difficult to characterize the environments of high-redshift quasars without access to the rest-frame optical at high sensitivity and resolution.

The study of high-redshift quasar environments was advanced significantly by the launch of JWST, which provides photometric bands extending much further longward of the Lyman break, and spectroscopy able to probe rest-optical emission lines, rather than Lyman-$\alpha$ alone, to spectroscopically confirm overdensities surrounding high-redshift quasars \citep{Champagne2025a, Kashino2023, Wang2023} and Little Red Dots \citep{Schindler2024}. While many studies with HST, ground-based telescopes, and now JWST have detected significant overdensities around targeted quasars, which may evolve into the massive galaxy clusters observed in the low-redshift Universe, some quasars do not have significant surrounding overdensities \citep[see e.g.][]{Banados2013, Eilers2024, RojasRuiz2024, Pudoka2025}, and instead live in underdense environments, or environments consistent with the field. 

A particularly interesting example of a quasar apparently without a surrounding overdensity is the $z=7.08$ quasar ULAS J112001.48+064124.3 (hereafter J1120+0641). Even among the (already extreme) population of bright high-redshift quasars, J1120+0641 is an outlier. At $z=7.08$, it was at the time of its discovery the highest-redshift quasar known \citep{Mortlock2011}, and measurements from single-epoch spectroscopy of various UV, optical, and near-IR lines place its black hole mass somewhere around $1-2\times10^9$ M$_{\odot}$, only about 750 Myr after the Big Bang. Its host galaxy is extremely bright in the far-infrared \citep{Venemans2020}, indicating significant dust-obscured star formation; JWST/NIRSpec IFU spectroscopy reveals a merging system of two galaxies \citep{Marshall2024}, and multi-band JWST/NIRCam photometric observations, with the bright quasar point-spread function (PSF) subtracted, point to a combined stellar mass of the merging components less than about $10^{10}$ $M_{\odot}$ \citep{Marshall2024, Stone2024}, corresponding to $M_{\mathrm{BH}}/M_* \sim 0.2$ even after the merger. This is further from the local $M_{\mathrm{BH}}/M_*$ relation than any other bright quasar at the end of the Epoch of Reionization yet studied \citep{YSun2025}. 

If J1120+0641 is to move towards the local $M_{\mathrm{BH}}/M_*$ relation, it must accrete significant mass from surrounding galaxies, and therefore must live in a dense environment. But here, too, J1120+0641 is apparently strange: \cite{Simpson2014} targeted J1120+0641 with deep HST ACS and WFC3 observations and, using a color-color selection technique to search for Lyman break galaxies, identified two $z\sim7.1$ galaxy candidates in their 13 arcmin$^2$ field. This is insufficient evidence for an overdensity. In fact, the presence of only two galaxies near the quasar redshift, in a 13 arcmin$^2$ field where $\sim5.8$ $z\sim7$ galaxies are predicted from the field luminosity function, suggested that J1120+0641 may live in a significantly {\em underdense} environment. With the HST data alone, there appears to be no way for the host galaxy of J1120+0641 to gain enough mass to bring its host galaxy in line with the mass of its black hole. 

However, \cite{Simpson2014} had access to only three HST photometric bands, with their longest (F125W) corresponding to rest-frame $\sim150$ nm at $z=7.08$. Is the nondetection of a galaxy overdensity around J1120+0641 simply a result of the limited wavelength coverage of the existing HST data? If re-observed with JWST, would a galaxy overdensity reveal itself and, furthermore, reveal enough mass to grow J1120+0641's host galaxy by the nearly two orders of magnitude necessary to bring it to the local local $M_{\mathrm{BH}}/M_*$ relation?

We designed a JWST GTO program (PID 2774) to re-observe the environment of the quasar J1120+0641. The quasar redshift places the [O~III] $\lambda$5007 emission line within the NIRCam F405N filter. At $z\sim7$, most galaxies should be vigorously forming stars and be bright in [O~III]: this narrow-band selection therefore provides a nearly complete sample of galaxies at or near the redshift of the quasar. By combining HST archival photometry and additional JWST broadband photometry with our narrow-band data, we probe the environment of J1120+0641 and examine future possibilities for this unique system. 

Our HST and JWST data and their reduction are described in Section \ref{sec:obs}. We outline the candidate selection and spectral energy distribution (SED) fitting process in Section \ref{sec:results}. Our determination of the environment of J1120+0641 is discussed in Section~ \ref{sec:environment}, and the implications of our results are explained in Section~\ref{sec:interpretation}. We conclude in Section \ref{sec:summary}. Throughout this work, we use a flat cosmology with H$_0$ = 69.6 km s$^{-1}$ Mpc$^{-1}$, $\Omega_m$ = 0.286, and $\Omega_\Lambda = 0.714$ \citep{Bennett2014}. All magnitudes reported are in the AB system.

\section{Observations and Data Reduction} \label{sec:obs}

\subsection{JWST NIRCam}

At $z=7.08$, the redshift of J1120+0641, the [O~III] 5007\angstrom\ emission line is shifted to 4.05 $\mu$m, into the NIRCam F405N narrow-band filter. We obtained deep observations in this filter ($t_{\mathrm{exp}}$ = 10522.0 s) and in the wider, overlapping F410M filter ($t_{\mathrm{exp}}$ = 5239.5 s) to search for [O~III]-emitting galaxies at the quasar redshift. The F360M filter ($t_{\mathrm{exp}}$ = 5239.5 s) provides a measurement of the adjacent continuum. 

We pair each of these NIRCam LW filters with a SW filter. F090W ($t_{\mathrm{exp}}$ = 10522.0 s) straddles the Lyman break at the quasar redshift, and together with the deep archival HST F814W observations (see Section \ref{sec:hst}) will add valuable Lyman-break redshift constraints to exclude lower- and higher-redshift interlopers. Of particular concern is H$\alpha$, which enters the F405N bandpass at $z\sim5.1$: these galaxies may be easily confused for $z\sim7.1$ [O~III] emitters. We therefore also observe in F150W ($t_{\mathrm{exp}}$ = 5239.5 s) and F200W ($t_{\mathrm{exp}}$ = 5239.5 s), which fall on either side of the Balmer break at $z\sim5.1$ and allow us to more easily identify and remove H$\alpha$ contaminants.

provide additional photometric redshift constraints to exclude lower- and higher-redshift interlopers from our $z=7.08$ sample. In particular, F090W straddles the Lyman break and F814W lies blueward of it for galaxies near $z=7.08$.

We observed the field using both NIRCam modules, with the quasar in Module B. To mitigate cosmic rays and detector artifacts, we adopted a $4\times1$ dithering pattern, using primary dither type INTRAMODULEBOX and subpixel dither type STANDARD and the DEEP2 readout pattern.

We processed our NIRCam data using the JWST pipeline version 1.14.1 \citep{Bushouse2023}, roughly following the procedures recommended in the STScI JWebbinars.\footnote{https://www.stsci.edu/jwst/science-execution/jwebbinars} The pipeline parameter reference file is registered in the JWST Calibration Reference Data System (CRDS) as jwst\_1235.pmap. We added a custom step to Stage 2 of the pipeline to characterize and subtract 1/f noise, a striping pattern in the images caused by the detector readout, from each frame. We produced final mosaic images in each band in Stage 3 of the pipeline by aligning and stacking all frames obtained in Stage 2, and resampled the images to a smaller pixel scale (0.0300\arcsec/pixel) using the drizzling algorithm in the Resample step of the pipeline. Finally, we used the Photutils Background2D function on the mosaiced and resampled images to perform a global background subtraction. We reach aperture-corrected 5$\sigma$ point-source depths of 0.16, 0.15, 0.16, 0.29, 1.72, and 0.42 $\mu$Jy in F090W, F150W, F200W, F360M, F405N, and F410M respectively, measured in apertures containing 80\% of the encircled energy for each PSF.

\subsection{HST} \label{sec:hst}

HST images in the F814W filter were taken with the ACS wide field camera between March 27 and March 29, 2013 as part of GO-13039 (PI: Simpson).  Observations were spread over 9 visits and 25 individual exposures, totaling 28,448 seconds. We re-reduced this data as follows.

Relative astrometry within a single visit was good, but small astrometric offsets existed between visits, which, if not corrected, produced artifacts in the final combined image. These were corrected by first combining the exposures taken in each visit, using \texttt{astrodrizzle}, then finding relative translational and rotational offsets between the combined, drizzled images using \texttt{tweakreg}.  Alignment was done using these drizzled images to preserve the relative astrometry between exposures in a given visit, and because combining multiple images increases effective depth and reduces the impact of cosmic rays, thus improving source detection and centroiding, ultimately improving the astrometric solutions.  Those solutions were propagated back to individual exposures using \texttt{tweakback}.  Finally, the 25 astrometrically-corrected frames were combined using \texttt{astrodrizzle} onto north-up grids with 0.0300\arcsec/pixel pixels (matching the JWST data), using a Gaussian kernel and a drop size (\textit{final\_pixfrac}) of 0.8.

\section{Analysis and Results} \label{sec:results}

To generate source positions and apertures to extract photometry, we use the MAST-generated F405N segmentation image, as we are only interested in sources detected in this band. 1054 sources are identified in the F405N image. 

We derive Kron radii, semi-major and semi-minor axes, and orientations for each source in the F405N segmentation image using {\tt photutils}, and use these parameters to extract aperture photometry in the remaining bands. Because the F405N photometry has a lower SNR than the other bands, we explored using the F410M segmentation image to derive aperture parameters: the resulting apertures were not systematically different than those determined from the F405N image.

We use the random apertures method to calculate errors on the measured fluxes. After masking out all sources in the image, we generate 2000 random positions within the image and place the aperture and measure photometry at each. Some fraction of these generated positions invariably fall on the gap between the NIRCam modules, so we set the value of these pixels to NaN so they will not affect the noise calculated.

\subsection{Candidate Identification}\label{sec:candidates}

We require that each galaxy be present (not necessarily significantly detected) in all bands from F150W to F410M (measured flux in aperture greater than zero, within the error bars\footnote{We ensured this did not artificially limit our completeness: relaxing this requirement adds 17 additional sources, almost all of which are spurious (see next paragraph).}). We expect the Lyman break to fall in the F090W filter, so we do not require that the source be present in this band (or F814W, when available). We are primarily interested in galaxies with a flux excess in the narrow band, F405N, compared to the adjacent ``continuum" band F360M, so we make the initial color cut $m_{\mathrm{F360M}} - m_{\mathrm{F405N}} > 0$. 454 of the 1054 sources fit these criteria.

However, members of a larger overdensity around J1120+0641 may have a larger velocity distribution, and their [O~III] emission may fall outside the very narrow F405N bandpass. We also therefore include galaxies that are not elevated in F405N relative to F360M, but {\em are} elevated in the overlapping medium band F410M ($m_{\mathrm{F360M}} - m_{\mathrm{F410M}} > 0$). This added 69 sources to our sample, for a total of 523 candidates.

\begin{figure*}
    \centering
    \includegraphics[width=16cm]{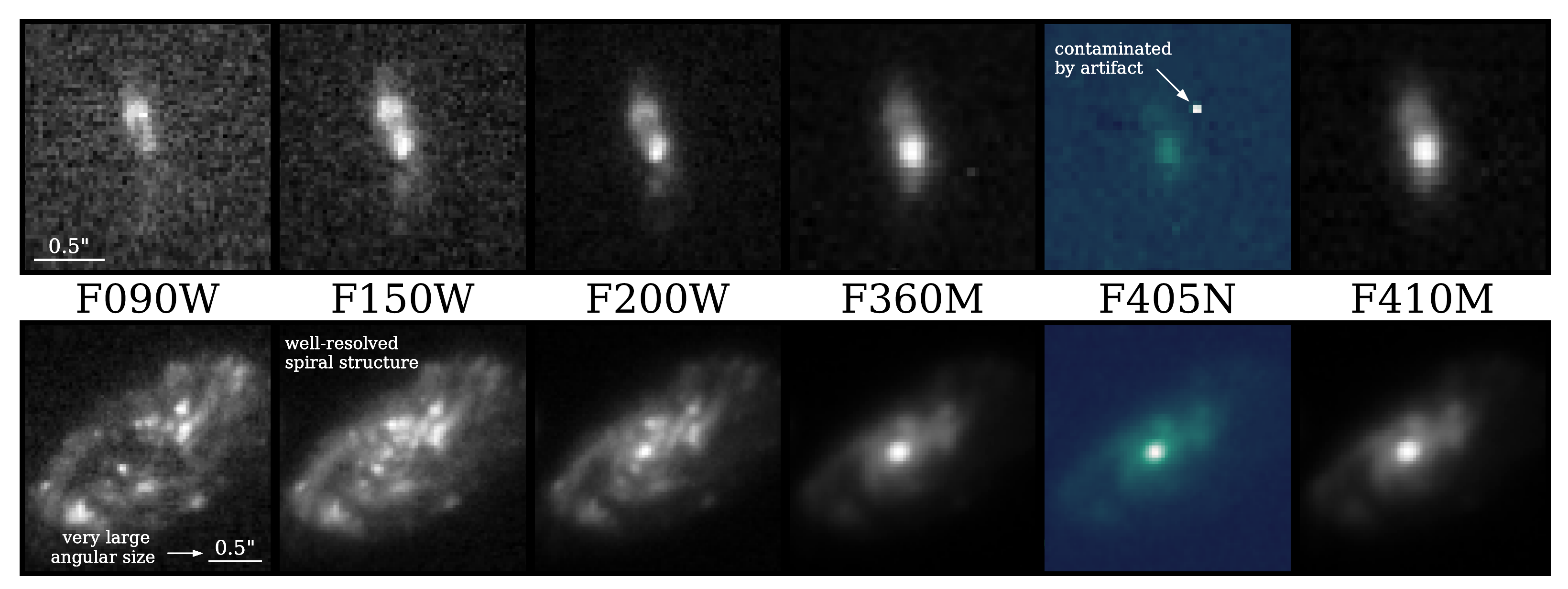}
    \caption{Examples of galaxies with elevated F405N or F410M that were removed from our sample in the visual inspection process. Candidate 048 {\it(top)} is classified as a spurious source: the underlying galaxy certainly exists, but its elevated narrow-band flux is likely due to the bright detector artifact visible in the F405N image. Candidate 163 {\it(bottom)} is classified as a low-redshift source based on its angular size, brightness, and clearly resolved spiral structure.}
    \label{fig:inspection}
\end{figure*}

We visually inspected these 523 candidates to narrow the sample  before performing SED fitting. 69 of the 523 sources are spurious: these are a mix of detector artifacts, shredded diffraction spikes of bright point sources in the image, and real galaxies only partially overlapping with the image in one or more bands (see the top panels of Figure \ref{fig:inspection} for an example). We also remove 112 galaxies obviously at very low redshift based on clearly resolved spiral structure/dust lanes (88 galaxies). Galaxies at $z > 3$ are very unlikely to have diameters (defined as twice  their effective radii) $> 1''$ \citep{Raikov2025}, so we impose a conservative cut at spatial extent greater than $ 2''$  (103 galaxies). Finally, we perform a flux cut to remove any with fluxes inconsistent with even the most extreme $z\sim7$ galaxy ($\gtrsim$1 $\mu$Jy in the UV continuum bands F090W, F150W, and/or F200W) not discarded in the visual inspection. This removes another 48 galaxies. For examples of discarded galaxies, see the lower panels of Figure \ref{fig:inspection}. These low-redshift galaxies are sufficiently different from possible contaminants in our study (in SED as well as morphology) that they do not need further study.

Our visually inspected sample consists of 294 candidates elevated in F405N and/or F410M relative to F360M, which we pass to SED fitting.

\subsection{Spectral Energy Distribution Fitting}

For two independent determinations of the exact redshifts of our 294 candidates, we pass them through two SED fitting codes: \texttt{Prospector} and \texttt{Bagpipes}.

We adopted a modified version of the \texttt{Prospector} code \citep{Johnson2021}, which is described in full by \cite{Lyu2024}. Briefly, we incorporate FSPS to model stellar emission, and model the AGN component and galaxy dust emission via semi-empirical templates. For the stellar component, we assumed a delayed-tau star formation history with a \cite{Kroupa2001} initial mass function, the \cite{Kriek2013} extinction law, and the associated nebular emission components, following the default parameter priors. Given the low-mass properties of these galaxies and the dominant obscured AGN population at high-$z$ \citep[e.g.,][]{Lyu2024}, we expect no significant AGN contribution to the rest-frame UV-optical SEDs of these galaxies, so the AGN component is turned off.  For the galaxy dust component, we adopted the Haro 11 template without the requirement of energy balance (as we lack constraining rest-IR data). 

We ran \texttt{Prospector} under these conditions three times for the 294 candidates identified in Section \ref{sec:candidates}, with different tophat redshift priors: $z=[0.1-13.0]$ (v1), $z=[5.0-9.0]$ (v2), and finally $z=[7.04-7.15]$ (v3; the [O~III] line shifts into the F405N bandpass for this redshift range). Galaxies far from the quasar redshift will be worse-fit under the stricter redshift priors, yielding a larger $\chi^2$. On the other hand, galaxies near the quasar redshift are likely to have a smaller $\chi^2$ value in v3 than in v2 or v1, and/or be fit to $z\sim7$ in the less restrictive v1 or v2 runs.

We identify 32 sources with a smaller $\chi^2$ with the most restrictive redshift prior (v3) than in v2 or v1. We also find 5 sources that have higher $\chi^2$ values in v3, but are also fit to $z=7-7.4$ in the lower $\chi^2$ v1 or v2 runs. Our \texttt{Prospector}-curated sample therefore consists of 37 galaxies total.

We also fit the 294 candidates with \texttt{Bagpipes}. We ran \texttt{Bagpipes} with a delayed-tau star formation history and \cite{Kroupa2001} initial mass function, as with \texttt{Prospector}. The \cite{Kriek2013} dust law is not supported in \texttt{Bagpipes}, so we use the \cite{Salim2018} law, which is parameterized similarly. We allow the maximum age of the stellar population to vary from zero to the age of the Universe at the fit redshift. The total mass of stars formed was allowed to vary from 10$^6$ to 10$^{12}$ M$_{\odot}$, and the metallicity from 0.1\% to 100\% solar. Modifying the functional form of the star formation history did not significantly affect the resulting stellar mass (5-10\% difference) or quality of the fit. For most sources, modifying the dust extinction law to a CF00 or Calzetti law changed the resulting stellar mass by $\lesssim20\%$, though in a few cases the mass changed by up to $\sim80\%$.  As with \texttt{Prospector}, we ran \texttt{Bagpipes} three times with three different redshift ranges: $z=[0.1-13.0]$ (v1), $z=[5.0-9.0]$ (v2), and finally $z=[7.0 - 7.5]$ (v3). The v3 range is larger for \texttt{Bagpipes} than for \texttt{Prospector}, because \texttt{Bagpipes} imposes a hard limit at the edges of the range, while \texttt{Prospector} can fit galaxies to a redshift outside the prior range.

\texttt{Bagpipes} fits 24 galaxies to a v3 $\chi^2$ lower than or comparable to v2 and v1; 22 of these candidates overlap with the \texttt{Prospector} sample. Of the 15 galaxies fit to $z\sim7.1$ in \texttt{Prospector} but not \texttt{Bagpipes}, nine are only mildly elevated in F405N relative to F360M and F410M, and \texttt{Bagpipes} fits them to $z\sim6.6$ based on the position of their Lyman break. Three more are fit to $z\sim5.1$, with F405N and/or F410M on H$\alpha$, and the remaining three are fit to $z<2$. These galaxies also tend to have larger $\chi^2$ values in the \texttt{Prospector} fits, and tend to lie at the bright end of the \texttt{Prospector} sample.

We therefore proceed with the sample of 22 galaxies which both \texttt{Prospector} and \texttt{Bagpipes} robustly fit to $z\sim7.1$. Their properties are summarized in Table \ref{tab:sample}. We note that the redshift range probed by the F405N filter, $z\in[7.04, 7.15]$, is not perfectly centered at the quasar redshift ($z=7.08$), which may introduce a sampling effect. We do not believe this makes a significant difference in our final sample, as the vast majority of our final sample have best-fit redshifts near the center of this range.

Of the 272 candidates not robustly localized to $z\sim7.1$, the majority ($\sim70\%$) fall into one of two categories. 118 candidates are fit by both \texttt{Prospector} and \texttt{Bagpipes} as Lyman-break galaxies at high redshift. These galaxies generally display a flat spectrum longward of F090W, and decline sharply in F090W before disappearing altogether in F814W. Another 50 galaxies are fit to $1.0\lesssim z \lesssim1.5$ by both codes. These galaxies tend to be very red across our observed bands, and are fit with older stellar populations (stellar age generally $\gtrsim 1$ Gyr) and $A_V \gtrsim 1$. These galaxies are bright in F360M, F405N, and F410M, which sit on the peak of the old stellar emission at 1.5-2 $\mu$m, and the flux declines at shorter wavelengths. This decline in the bluer bands can mimic a Lyman break: a further 18 galaxies are fit to $1.0\lesssim z \lesssim1.5$ by \texttt{Bagpipes} but with a Lyman break ($z\gtrsim8$) by \texttt{Prospector}.

\begin{figure*}
    \centering
    \includegraphics[width=18cm]{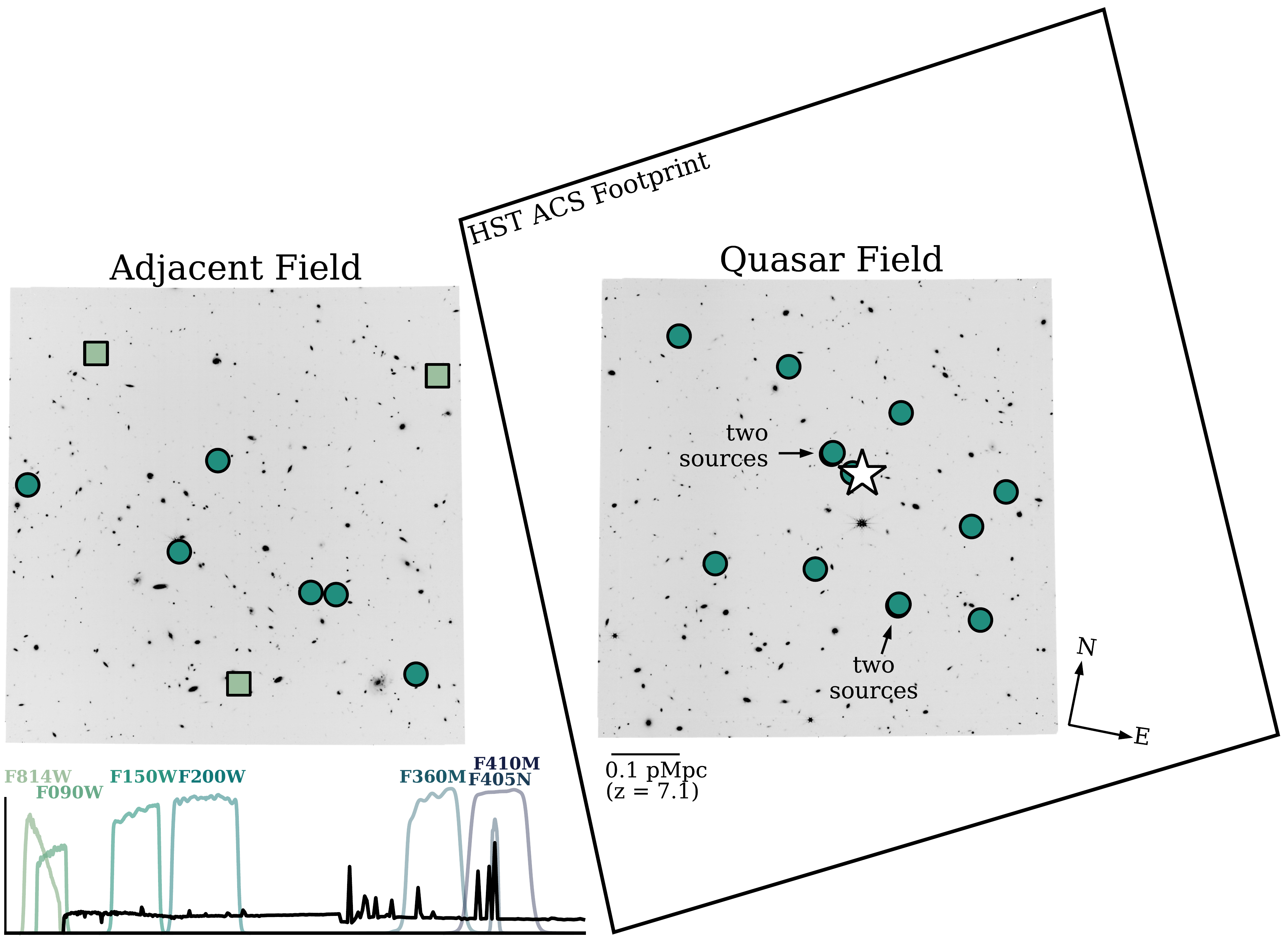}
    \caption{The field around J1120+0641, and the narrow-band candidates identified. The surveyed JWST field in F410M is shown in black and white, with $z\sim7$ candidates marked with green circles. The footprint of the archival HST F814W observations is also shown, centered on the quasar position, which is marked with a white star. Some sources in the quasar field overlap one another, and are indicated (``two sources"). The three candidates in the adjacent field with less certain redshift determinations are marked with lighter green squares. At lower left, we show the transmission curves of the HST/ACS and JWST/NIRCam filters observed, along with an example $z=7.1$ spectrum. The [O~III] $\lambda5007$\angstrom $\,$ line shifts into the NIRCam F405N filter for a narrow redshift range around $z\sim7.1$.}
    \label{fig:candidates}
\end{figure*}

\subsection{Final $z\sim7$ Sample}\label{sec:sample}

\begin{deluxetable*}{c|ccc|ccccc}
\tablecaption{Properties of our $z\sim7.1$ sample. The location of the candidate in the quasar (Q) or adjacent (A) field is indicated in Column 4. We report the best-fit redshifts and stellar masses from \texttt{Prospector} and \texttt{Bagpipes} in Columns 6-9. The three medium-confidence candidates are listed at the bottom with an asterisk. \label{tab:sample}}
\tablehead{
\colhead{ID} & \colhead{RA ($^\circ$)} & \colhead{Dec ($^\circ$)} & \colhead{Field} &  \colhead{$m_{\mathrm{F150W}}$} & \colhead{$z_{\mathrm{ps}}$\tablenotemark{a}} & \colhead{$z_{\mathrm{bp}}$\tablenotemark{a}} & \colhead{$\log\,(M^*_{\mathrm{ps}}$/M$_{\odot}$)} & \colhead{$\log\,(M^*_{\mathrm{bp}}$/M$_{\odot}$)}
}
\decimalcolnumbers
\startdata
300 & 169.99902 & 6.70370 & Q & $29.03 \pm 0.41$ & 7.11 & 7.10 & $8.36\pm0.37$ & $7.43\pm0.29$ \\
335 & 169.99734 & 6.69700 & Q & $26.64 \pm 0.21$ & 7.10 & 7.11 & $8.98\pm0.19$ & $8.51\pm0.27$ \\
337 & 169.99750 & 6.69704 & Q & $27.21 \pm 0.34$ & 7.11 & 7.11 & $9.02\pm0.51$ & $7.89\pm0.20$ \\
414 & 169.99737 & 6.68958 & Q & $29.20 \pm 0.69$ & 7.10 & 7.12 & $8.71\pm0.52$ & $8.15\pm0.61$ \\
427 & 169.99444 & 6.68186 & Q & $27.70 \pm 0.21$ & 7.11 & 7.11 & $8.66\pm0.29$ & $7.84\pm0.23$ \\
534 & 170.00581 & 6.69992 & Q & $27.58 \pm 0.33$ & 7.11 & 7.11 & $9.12\pm0.41$ & $8.10\pm0.25$ \\
624 & 170.00959 & 6.70138 & Q & $28.51 \pm 0.31$ & 7.11 & 7.11 & $8.56\pm0.29$ & $7.71\pm0.31$ \\
666 & 170.00589 & 6.68920 & Q & $27.99 \pm 0.77$ & 7.11 & 7.11 & $9.19\pm0.76$ & $8.73\pm0.66$ \\
703 & 170.00662 & 6.68699 & Q & $27.20 \pm 0.27$ & 7.11 & 7.11 & $9.21\pm0.45$ & $7.89pm0.18$ \\
709 & 170.00678 & 6.68705 & Q & $28.74 \pm 0.77$ & 7.07 & 7.10 & $9.26\pm0.69$ & $8.43\pm0.74$ \\
789 & 170.01207 & 6.69086 & Q & $28.03 \pm 0.37$ & 7.14 & 7.11 & $9.02\pm0.56$ & $8.03\pm0.42$ \\
876 & 170.01181 & 6.68086 & Q & $27.80 \pm 0.31$ & 7.10 & 7.11 & $8.37\pm0.36$ & $7.60\pm0.23$ \\
941 & 170.01047 & 6.67159 & Q & $28.02 \pm 0.40$ & 7.13 & 7.11 & $9.00\pm0.45$ & $8.05\pm0.17$\\
\hline
181 & 169.97607 & 6.66299 & A & $27.85 \pm 0.38$ & 7.11 & 7.08 & $9.12\pm0.42$ & $8.08\pm0.35$ \\
358 & 169.97943 & 6.65433 & A & $26.90 \pm 0.17$ & 7.11 & 7.08 & $8.96\pm0.27$ & $8.41\pm0.15$ \\
366 & 169.97875 & 6.65236 & A & $27.76 \pm 0.35$ & 7.11 & 7.08 & $9.34\pm0.45$ & $7.96\pm0.29$ \\
456 & 169.97744 & 6.64111 & A & $26.05 \pm 0.11$ & 7.12 & 7.08 & $8.83\pm0.30$ & $8.16\pm0.10$ \\
634 & 169.97744 & 6.62748 & A & $27.47 \pm 0.37$ & 7.09 & 7.08 & $9.48\pm0.55$ & $8.25\pm0.31$ \\
693 & 169.98564 & 6.64099 & A & $27.48 \pm 0.21$ & 7.11 & 7.08 & $8.89\pm0.24$ & $8.07\pm0.23$ \\
160* & 169.96941 & 6.64997 & A & $29.50 \pm 0.72$ & 7.23 & 7.31 & $10.41\pm0.25$ & $9.26\pm0.25$ \\
856* & 169.99940 & 6.65472 & A & $28.04 \pm 0.26$ & 7.01 & 7.41 & $11.25\pm0.23$ & $10.12\pm0.30$ \\
911* & 169.98969 & 6.62826 & A & $27.50 \pm 0.16$ & 7.20 & 7.17 & $11.29\pm0.25$ & $9.60\pm0.22$ \\
\enddata
\tablenotetext{a}{The uncertainties on the redshifts are small due to the narrow-band selection method; the presence of the bright F405N and the Lyman break in F090W and F814W robustly places the high-confidence candidates between $7.04\leq z \leq7.15$. The position of the Lyman break is not as clear for the three medium-confidence candidates (see Figure \ref{fig:appendix3}), and they are generally not as bright in F405N relative to F360M. Their redshift errors are slightly larger, closer to $\pm0.2$. }
\end{deluxetable*}

The positions of the 22 $z\sim7.1$ candidates are shown, along with the position of the quasar, in Figure \ref{fig:candidates}, and their thumbnails and SEDs are plotted in the Appendix. Thirteen candidates lie in Module B, in the same module as the quasar; we hereafter refer to Module B as the ``quasar field." The remaining nine candidates lie in Module A, hereafter the ``adjacent field" (though note that an overdensity around J1120+0641 is likely to span both fields). The vast majority of these candidates are fit to a redshift between 7.07 and 7.14 by both \texttt{Prospector} and \texttt{Bagpipes}, but three of the candidates in the adjacent field are fit farther from the quasar redshift by both codes. We consider these three candidates ``medium confidence," and they are shown in a lighter color and with square markers in Figure \ref{fig:candidates}.

Approximately a quarter of this sample displays multi-component morphologies (see cutouts in Appendix). To ensure that none of these sources were chance alignments of galaxies at two different redshifts, we performed deblended photometry on this subset of the sample using smaller, hand-placed apertures. In all cases, the two components displayed nearly identical SED shapes. It is therefore unlikely that any of these ``clumpy" sources are due to projection effects.

Despite the similar model parameters input to both codes, the stellar masses returned by \texttt{Prospector} are larger on average than those from \texttt{Bagpipes}, by approximately an order of magnitude in most cases. This is mainly a consequence of our lack of data longward of rest-frame 500 nm. \texttt{Bagpipes} prefers to fit a very young stellar population while \texttt{Prospector} tends to fit an older stellar population. The two fits diverge beyond rest-frame 1 $\mu$m, but without longer-wavelength data to constrain the shape of the SED we cannot rule out either possibility. \texttt{Bagpipes} fits the Lyman-alpha break with a damping wing, while in some cases \texttt{Prospector} prefers Lyman-alpha emission and additional dust extinction in the UV (see figures in the Appendix). Both codes return similar relative $A_V$ between sources, though the Prospector values are shifted slightly higher overall (by $\sim40$\% on average, due to fitting the rest-UV with more dust). The best-fit star formation rates (SFRs) from both codes agree roughly with the SFR estimated from the F405N - F360M flux, i.e. from the [O~III] luminosity. For our purposes, the most important parameter is the stellar mass: we discuss the effect of the two codes' mass discrepancy on our results in Section \ref{sec:future}.

\section{The Environment of J1120+0641} \label{sec:environment}

\subsection{HST-only Candidates}

\begin{figure*}
    \centering
    \includegraphics[width=15cm]{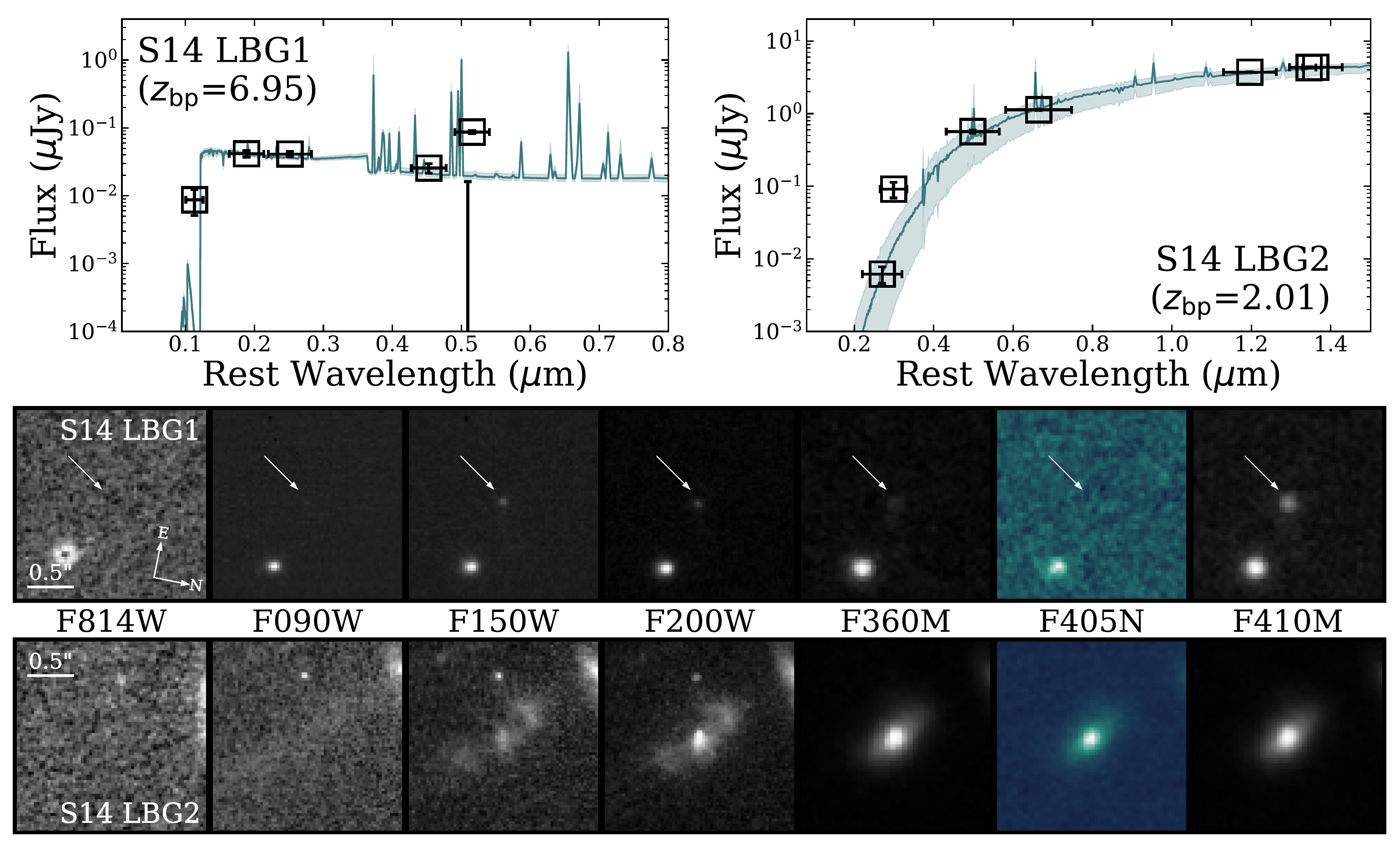}
    \caption{The SEDs and thumbnail images of the two \cite{Simpson2014} LBG candidates that overlap with our JWST data. LBG 1 (left SED, upper row of thumbnails) is not picked up in our sample due to its nondetection in F405N. Its SED shape and flux place it at $z\sim7$, but it is likely not associated with the quasar directly. LBG 2 (right SED, lower row of thumbnails) is exceedingly bright in the longer NIRCam bands, in agreement with its detection with Spitzer. Like \cite{Simpson2014}, we believe this galaxy likely lies at $z\sim2$.}
    \label{fig:hstonly}
\end{figure*}

\cite{Simpson2014}, using HST F814W, F105W, and F125W photometry as well as Subaru $i$ and $z$ band and Spitzer/IRAC images, identified three $z\sim7.1$ Lyman-break candidates, which they call LBG1, LBG2, and LBG3. LBG3 lies just outside the footprint of our JWST images, but LBG1 and LBG2 are included in our data. The SEDs and thumbnails of LBG1 and LBG2 are shown in Figure \ref{fig:hstonly}.

LBG1 is not detected in the F405N segmentation map, meaning it is not included in our narrow band-selected sample and its redshift determination is less certain. Its observed SED shape is consistent with a $z\sim7$ galaxy (see Figure \ref{fig:hstonly}), and \texttt{Bagpipes} fits it at $z=6.95$, primarily from the position of the Lyman break; given the width of the F090W bandpass, the uncertainty on this redshift is of order $\pm0.5$. This is in contrast to the high-confidence overdensity candidates, where the F405N excess combined with the Lyman break allows them to be fit to $7.08\leq z \leq 7.12$ with much smaller redshift uncertainties. LBG1 is certainly not a low-$z$ contaminant, but without an F405N excess we cannot confirm whether it is directly associated with the quasar and simply [O~III]-faint, or an unrelated galaxy at slightly lower or higher redshift than the quasar. Because of its uncertain redshift, we do not include LBG1 in the $z\sim7.1$ candidate sample going forward.

LBG2 is robustly detected by Spitzer at 3.6 and 4.5 $\mu$m. \cite{Simpson2014} fit the SED of LBG2, including the Spitzer data, with EAzY, which indicated LBG2 likely lies at $z\sim2$ rather than $z\sim7.1$. Our JWST data shows a steeply rising SED between F814W and F410M, unlike the rest of our $z\sim7.1$ candidates. LBG2 is extremely bright in the redder NIRCam bands ($>4 \mu$Jy in F405N and F410M, see top right of Figure \ref{fig:hstonly}), consistent with its Spitzer detections (though the IRAC1 flux of LBG1 is $\sim1.3\sigma$ higher than our F360M flux).  Because of this very red SED, LBG2 {\em is} selected by our narrow-band excess criterion, but is removed from our final candidate sample at the visual inspection stage because of its large flux and extended shape in the redder NIRCam filters. \texttt{Bagpipes} places LBG2 at $z=2.01$.

\subsection{Number Counts}\label{sec:counts}

\begin{figure}
    \centering
    \includegraphics[width=8cm]{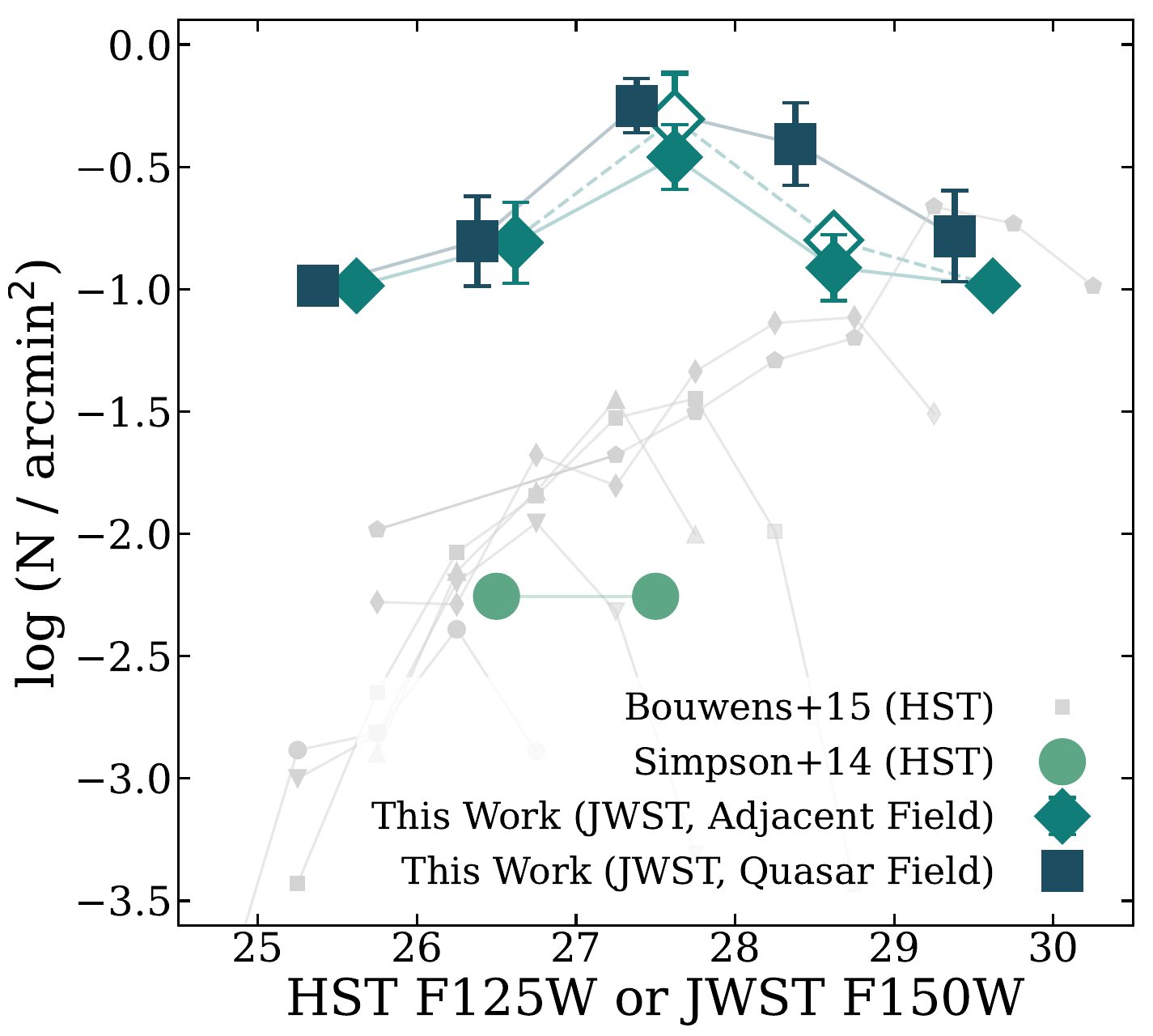}
    \caption{$z\sim7.05 - 7.15$ JWST galaxy number counts around J1120+0641 (navy squares) and in the adjacent field (teal diamonds). For the adjacent field, show counts incorporating the three medium-confidence candidates as open diamonds. For comparison, we plot very deep $z\sim7$ HST $J_{125}$ counts in five legacy fields from \cite{Bouwens2015} in grey, scaled to only reflect galaxies between $z=7.05$ and $7.15$ (see Section \ref{sec:counts} for details). Number counts derived from the two \cite{Simpson2014} candidates are shown as light green circles. With only HST data, the quasar field appears significantly underdense, but incorporating JWST data reveals a significant overdensity of galaxies at $m_{\mathrm{F150W}} \lesssim 28$, beyond which the lack of completeness at the faint end of our sample becomes evident. Markers have been offset horizontally for visual clarity. Counts and errors are calculated via a Monte Carlo method; see Section \ref{sec:counts} for details.}
    \label{fig:counts}
\end{figure}

We derive number counts via a Monte Carlo process. This allows us to take into account the uncertainty on source magnitudes, as each galaxy's uncertainty may shift it into a different magnitude bin. For each galaxy in the sample, we select a random value from a Gaussian centered at the object's F150W flux, with standard deviation set to the F150W uncertainty from random apertures. We convert these resampled fluxes to apparent magnitudes, then sort the galaxies into magnitude bins. By repeating this process 1000 times, we derive a distribution of the number of galaxies in each bin. We take the mean of this distribution as the final count in the bin, and the standard deviation as its uncertainty. 

The final counts in the quasar field (filled navy squares) and adjacent field (filled teal diamonds) are shown in Figure \ref{fig:counts}. For the adjacent field, we also include the counts derived when we include the three medium-confidence candidates (open teal diamonds).

Comparing these counts to \cite{Simpson2014} is not trivial. Plotting the raw number of candidates detected will not properly represent the difference in sensitivity between our two samples, because the redshift range probed by our narrow-band selection method is very different from that probed by their three-band color-color technique. In theory, [O~III] selection with the F405N filter should probe galaxies between $z\sim 7.05$ and $7.15$, which agrees with our best-fit redshifts from SED fitting (see Table \ref{tab:sample}). A color-color selection probes a much larger redshift range, i.e. if all else is equal (depth, etc.), their technique will select more galaxies per unit area.

Therefore, to robustly compare our counts with those from \cite{Simpson2014}, we need to estimate the redshift range probed by their LBG selection technique, and calculate the fraction of that range that will lie within $z \in [7.05, 7.15]$. \cite{Simpson2014} verified their selection method by degrading images of the Hubble Ultra Deep Field (HUDF) to the depth of their quasar observations. They used their color-color technique to select a sample of $z\sim7$ galaxies in the HUDF, and compared it to canonical HUDF $z\sim7$ samples derived with the full filter set and depth of the HUDF. Their selection picks up five HUDF galaxies, which have robust photometric redshifts from \cite{McLure2013} between $z=6.5$ and $7.7$. We therefore assume that the \cite{Simpson2014} color-color LBG selection technique uniformly selects galaxies between $z=6.5$ and $7.7$, meaning that 1/12th of galaxies selected with their method will lie between $z=7.05$ and $7.15$.  

For this reason, we scale down the counts from the two \cite{Simpson2014} sources by a factor of 12 and plot them in Figure \ref{fig:counts} as teal circles; we do not perform the Monte Carlo procedure, and so do not include uncertainties on these counts.

We also plot HST $z\sim7$ counts in multiple fields from \cite{Bouwens2015}, the deepest of which reaches a 5$\sigma$ depth of $\sim30$ mag. Like the \cite{Simpson2014} data, these counts probe a much larger redshift range than our narrow-band selection technique. However, \cite{Bouwens2015} provide the photometric redshift distribution of their $z\sim7$ sample (their Figure 1), which allows us to determine that approximately $5\%$ of their sample should lie within $z \in [7.05, 7.15]$. We scale their data accordingly in Figure \ref{fig:counts}.

With the \cite{Simpson2014} data alone, the quasar field appears slightly underdense compared to the field. However, with the addition of the JWST narrow-band selected galaxies, it is clear in Figure \ref{fig:counts} that both the quasar field and the adjacent field are significantly overdense in $z \in [7.05, 7.15]$ galaxies. We note that we are complete only to $\sim28$ mag F150W, which is responsible for the falling counts at fainter magnitudes. The counts in the two fields are similar down to the completeness limit, but the quasar field is slightly richer than the adjacent field. Incorporating the three medium-confidence candidates in the adjacent field does not affect this conclusion significantly.

\section{Interpretation}
\label{sec:interpretation}

ULAS J1120+0641 is one of the most thoroughly studied very high redshift quasars. We review the properties of the quasar and its present host galaxy in this section. 

\subsection{Black Hole Mass}
\label{sec:bhmass}

No dynamical mass measurement is available for the SMBH in J1120+0641, but the many single-epoch measurements of its mass are listed in Table \ref{tab:bhmass}, based on line widths and the standard single-epoch mass determination \citep[e.g.,][]{Vestergaard2006}. In general, this suite of measurements shows good agreement and indicates $M_{BH} \gtrsim 10^9$ M$_\odot$. The single-epoch technique based on line widths can be subject to systematic errors due, for example, to outflows, so we include two other measures, both of which agree with the conclusion from the conventional single-epoch determinations. 

The luminosity and temperature of the accretion disk are controlled by the black hole mass; models of the disk provide a means to estimate this mass \citep{Calderone2013}. \citet{Campitiello2019} used this method to determine possible masses for the black hole in J1120+0641. In Table~\ref{tab:bhmass} we show the range of values for different viewing angles, black hole spins, and using both the  relativistic model KERRBB \citep{Li2005} and a slim disk model \citep[e.g.,][]{Sadowski2009}.

Recently, \cite{King2024} suggested that many high-redshift quasars might have beamed outputs. This would raise their black hole signatures well above the true, unbeamed values, and the black hole masses calculated from observations of beamed quasars will be overestimated. We can evaluate the possibility of beamed emission from J1120+0641 from the velocities provided in \cite{Venemans2017, Bosman2024}; and \cite{Marshall2024}. From the latter reference we obtain a host [O~III] redshift of $7.0804 \pm 0.0028$ while from the first reference we find a host redshift from [C~II] of $7.0851 \pm 0.005$ (average 7.08275; weighted average 7.08496). The two broad components of H$\alpha$ have redshifts of 7.075 and 7.0936, corresponding to outflows at -315 and 360 km s$^{-1}$ respectively \citep{Bosman2024}. Taking the broad-line redshift to be the average and comparing with the host galaxy redshift, the galaxy velocity relative to the broad-line region is $<$ 100 km s$^{-1}$. This relatively small velocity difference is inconsistent with beaming: we therefore find no evidence that the single-epoch measurements of J1120+0641's black hole mass are overestimated. 

J1120+0641 has been detected in X-rays by Chandra and XMM-Newton \citep{Page2014, Moretti2014}. These two papers agree that the $\alpha_{OX}$ slope is consistent with that of typical lower-redshift X-ray quasars, i.e. that the X-ray luminosity is ``normal.'' The slope of the X-ray spectrum depends on the Eddington ratio \citep{Moretti2014}, and can provide another constraint on the black hole mass. We constrain the X-ray slope using the fractional 2 - 10 keV luminosity. The softening of the slope of the spectrum reduces the fractional luminosity (or, equivalently, increases the bolometric correction $L_{bol}/L_{[2-10kev]}$). \cite{Moretti2014} report $L_{bol}/L_{[2-10kev]} \sim 406^{+21}_{-17}$ for J1120+0641, taking the bolometric luminosity of $2.7 \times 10^{47}$ erg s$^{-1}$ (see  Appendix B; these errors only account for luminosity uncertainties). \citet{Page2014} find $L_{bol}/L_{[2-10kev]}$ of either $146^{+96}_{-53}$ or $574^{+137}_{-91}$ depending on whether the Chandra or XMM-Newton data is used to determine the 2 - 10 keV luminosity.

These $L_{bol}/L_{[2-10kev]}$ estimates are widely spread, but nonetheless indicate that the luminosity is not strongly super-Eddington. The theoretical models of \cite{Pacucci2024} show that Eddington ratios between 1 and $\sim$ 2.5 lead to $L_{bol}/L_{[2-10kev]} > 1000$ for $z\sim 6$ quasars, much higher than indicated for J1120+0641. Their models yield lower $L_{bol}/L_{[2-10kev]}$ for high Eddington ratios only if the black hole is viewed pole-on and has significant spin, but these conditions are unlikely for J1120+0641 given the lack of beaming. This implies that J1120+0641 is not accreting significantly above the Eddington limit. 

Similarly, the models of \citet{Inayoshi2024} (see their Figure 4) place J1120+0641 below unity Eddington ratio (i.e., $\lambda = 1$). Allowing for errors, a highly super-Eddington ratio is not plausible, although a mildly super-Eddington one is possible.

Higher super-Eddington ratios are also unlikely for J1120+0641 because they imply a black hole mass well below the broad line estimates.  
From these considerations, we place an upper limit on the Eddington ratio, $\lambda < 2$. Since the luminosity is well determined (Appendix B), this puts a lower limit on the black hole mass as indicated in Table~\ref{tab:bhmass}. Combining all the methods, we adopt a fiducial value of $M_{BH} = 1.5 \times 10^9$ M$_\odot$.

\begin{deluxetable}{cccc}
\tablecaption{Selected SMBH mass measurements for ULAS J1120}
\label{tab:bhmass}
\tablehead{
\colhead{} & \colhead{$M_{\mathrm{BH}}$ ($10^9$ M$_{\odot}$)} & \colhead{Error} & \colhead{Reference}
}
\startdata
\hline
  \multicolumn{4}{c}{Broad Lines (Single-Epoch)}  \\
  \hline
   MgII & 2.4 & $0.2$ &  \citet{Derosa2014} \\
  CIV & 1.09 & $^{+0.02}_{-0.04}$ &  \citet{Derosa2014} \\
  MgII \& CIV &  1.35  & 0.04  &  \citet{Yang2021}  \\
   CIV &  2.40  & $^{+0.06}_{-0.05}$  &  \citet{Farina2022}  \\
H$\alpha$ & 1.55 & 0.22 &  \citet{Bosman2024} \\
Pa$\alpha$  &    0.98 &  $^{+0.28}_{-0.23} $ & \citet{Bosman2024} \\
Pa$\beta$  &    0.87 &  $^{+0.21}_{-0.17}$ &  \citet{Bosman2024} \\
H$\beta$  & 1.4  & $^{+2.6}_{-1.1}$  & \citet{Marshall2024} \\
\hline\hline
 \multicolumn{4}{c}{Accretion Disk Modeling} \\
   &   1.3 - 7.7 &  -- & \citet{Campitiello2019} \\
\hline
\hline
  \multicolumn{4}{c}{Eddington Ratio $<$ 2}\\
   &   $\gtrsim$ 1  &  & this work \\
\enddata
\end{deluxetable}

\subsection{Host Galaxy Mass}

Our estimate of the mass of the galaxy  depends both on our understanding of the relevant observational parameters, and on the stellar initial mass function assumed. To put together a complete picture of the host galaxy's characteristics, we will discuss (1) the implications from the ULIRG-level SFR on the derived mass; (2) the mass of the stellar population seen in the UV - IR; and (3) the possible effects of the initial mass function (IMF). Given that J1120+0641 is already believed to have have a relatively low stellar mass relative to its $M_{\mathrm{BH}}$, we will also address deriving an upper limit to the stellar mass. 

\subsubsection{Far infrared emission and associated SFR}\label{sec:ulirg}

The J1120+0641 host has a substantial far-infrared (far-IR) luminosity, indicating a very high SFR. \cite{Venemans2020} use a rest-frame 1900 GHz (158 $\mu$m) continuum measurement to derive a far-IR luminosity, assuming a modified blackbody model for the far-IR SED, with a temperature of 47K and $\beta$ = 1.6.  They multiplied this FIR estimate by 1.41 to correct it to total infrared (TIR) luminosity. However, a modified blackbody may not be the best model at such high redshifts, and the FIR SED of low-redshift, low-metallicity Lyman continuum emitter Haro 11 should provide a more accurate model for the host \citep[see e.g.][]{Lyu2016, Derossi2018}. We therefore normalize the SED of Haro 11 to the 158 $\mu$m measurement and correct to TIR by integrating, finding a correction of a factor of 1.63 (rather than 1.41) and a ULIRG-level TIR luminosity of $1.9 \times 10^{12}$ L$_\odot$, implying an obscured SFR of 280 M$_\odot$ yr$^{-1}$ \citep{Kennicutt2012, Murphy2011, Hao2011}. 

\begin{figure}
    \centering
    \includegraphics[width=7.5cm]{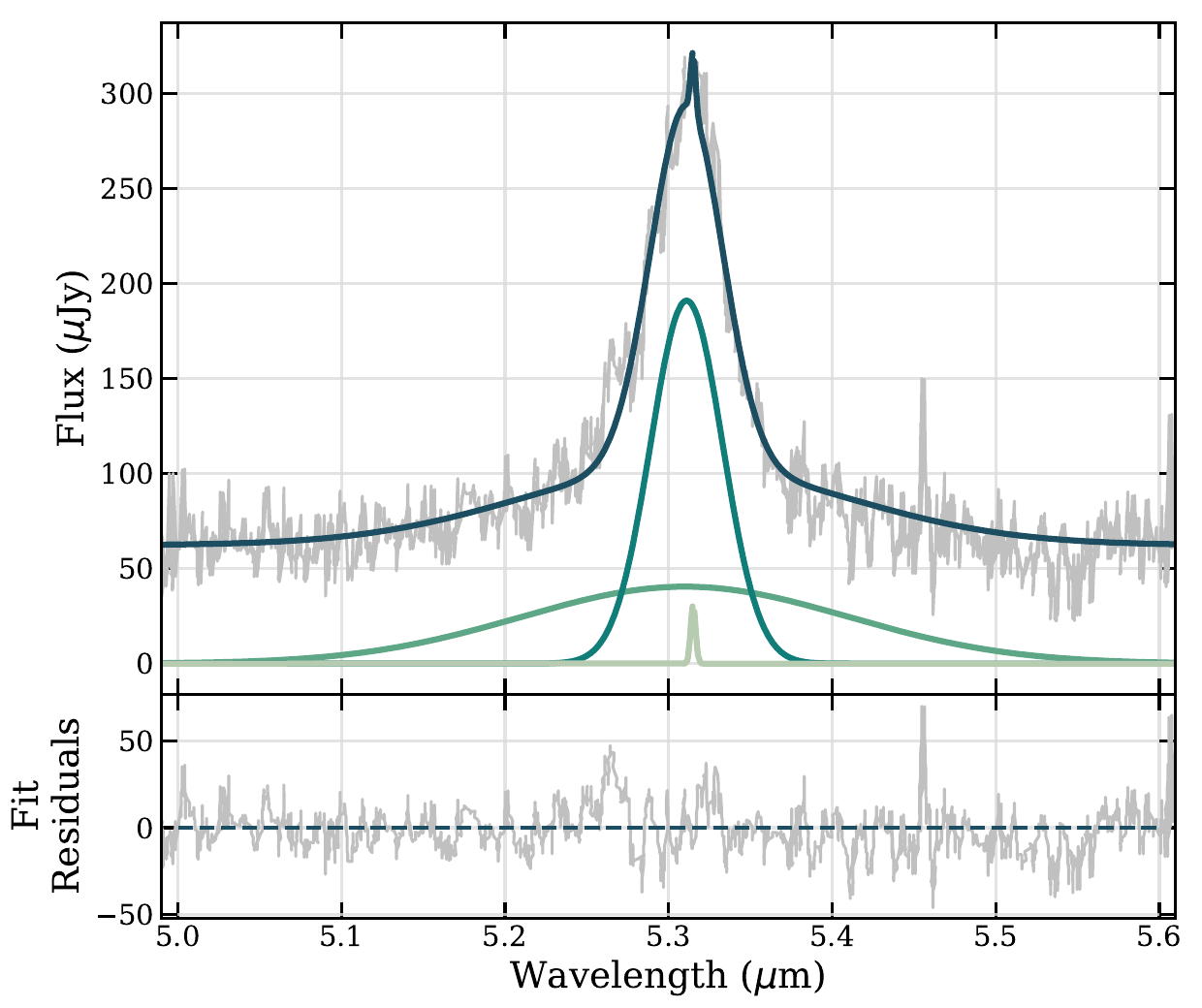}
    \caption{Three-line fit to the profile of the H$\alpha$ line. The two broad components and flat continuum are as in \citet{Bosman2024}, where we also obtained the data; we have added a faint unresolved line contributed by the star formation, which reduces the $\chi^2$ of the fit. The residuals are shown in the lower panel. }
    \label{fig:halph}
\end{figure}

In general, ULIRGs absorb most of the energy in emission lines due to young stars, with only a fraction that escapes, meaning SFR estimates from optical and near-IR emission lines underestimate the total star formation. We can check for this behavior by comparing J1120+0641's SFR from the infrared luminosity with what would be deduced from standard optical emission line estimates. The much lower SFR from H$\beta$ \citep[12.5 M$_\odot$ yr$^{-1}$,][]{Marshall2024} seems to support this point, but because of artifacts from their PSF subtraction, the region within a radius of $\sim$ 0\farcs2 is not included in this measurement. We expect this region to contain most of the central starburst, based on the 1 - 1.5 kpc  spatial extent of the far-IR emission \citep{Venemans2017}.  \citet{Bosman2024} measured H$\alpha$ in an aperture of radius 0\farcs3 centered on the nucleus, providing a possibility to fill in the missing area. In Figure~\ref{fig:halph}, we show the result of refitting their line profile, including two broad components as in their fit but adding an unresolved line. This unresolved component reduces the $\chi^2$ of the fit, based on an error estimated by repeating the fit multiple times with the narrow component set to different wavelengths. This fit is offset to the red by $\sim$ 200 km s$^{-1}$, consistent with the result in Section~\ref{sec:bhmass} given that we are probing different regions in the galaxy. The estimated flux of the narrow H$\alpha$ line is $1.2 \pm 0.5 \times 10^{-20}$ W m$^{-2}$. A similar fit to the Pa$\alpha$ profile, keeping the line widths and relative wavelengths (in velocity space) shows a weak detection ($\sim 1.5 \sigma$). The fact that neither H$\alpha$ nor Pa$\alpha$ show a strong narrow line supports the host galaxy's ULIRG behavior, indicating strong obscuration that suppresses optical/near-IR star formation signatures.

 The SFR deduced from the H$\alpha$ is about 40 M$_\odot$ yr$^{-1}$. Combining it with the \cite{Marshall2024} H$\beta$ measurement at larger radii yields a total of $\sim50$ M$_\odot$ yr$^{-1}$, less than 20\% of the FIR SFR. This difference must be attributed to very strong extinction in the star-forming region, supported by the compactness and strength of the far-IR emission. Such behavior is typical of high redshift ULIRGs \citep{Swinbank2004, Takata2006}.
 
The host galaxy of J1120+0641 therefore appears very similar to lower redshift ULIRGs of similar luminosity. The typical timescale for this level of star forming activity is less than 10 Myr \citep[e.g.,][]{Thornley2000, Bekki2006, Kawaguchi2020}. Thus the current burst of star formation can be expected to add about $1 - 2 \times 10^9$ M$_\odot$ to the host galaxy mass. These newly forming stars are hidden by the dust extinction from the rest optical and UV. We can therefore use existing HST and JWST observations of the quasar to characterize the older stellar population that dominates the mass.

\subsubsection{Older, optically visible host galaxy}\label{sec:galaxy}

A nominal mass for the host galaxy stellar population of $3.8 \pm 1.0 \times 10^9$ M$_\odot$ is reported in \cite{YSun2025}.  This mass was determined by comparison with field galaxies of similar mass and redshift, as originally described in \citet{Stone2024}, and is based on models using a Chabrier or Kroupa initial mass function. \citet{Marshall2024} estimated the mass using \texttt{Prospector} and a delayed-tau star formation history (with an optional late starburst). They obtained a mass of $2.6^{+2.4}_{-1.4} \times 10^9$ M$_\odot$. These estimates are in agreement, despite the different approaches. We adopt the average of the two, $3.2 \times 10^9$ M$_\odot$, as our best estimate. 

\citet{Stone2024} based their estimate on values derived from modeling field galaxies, rejecting those dominanted by very young stellar populations. We show in Section~\ref{sec:future} that \texttt{Prospector} fits such as those used by \citet{Marshall2024} also return masses assuming a mature stellar population. As documented in Section~\ref{sec:future}, if instead the host galaxy is dominated by very young stars, then its mass could be significantly lower than this estimate.

Our mass estimate  is unlikely to be ``missing" mass due to a compact stellar core behind the quasar or uncorrected dust extinction. \citet{Marshall2024} found the host to be well resolved, with an effective radius r$_{eff}$ = 0\farcs361 $\pm$ 0\farcs001. Their images of the host display virtually identical morphologies in F115W, F200W, and F356W, with half widths at half maximum of the diffraction-limited images ranging from 0\farcs02 to 0\farcs06. The presence of a compact central component that could add significantly to the total galaxy flux would require a major departure from a S\'{e}rsic profile and seems quite unlikely. 

\begin{figure}
    \centering
    \includegraphics[width=7.5cm]{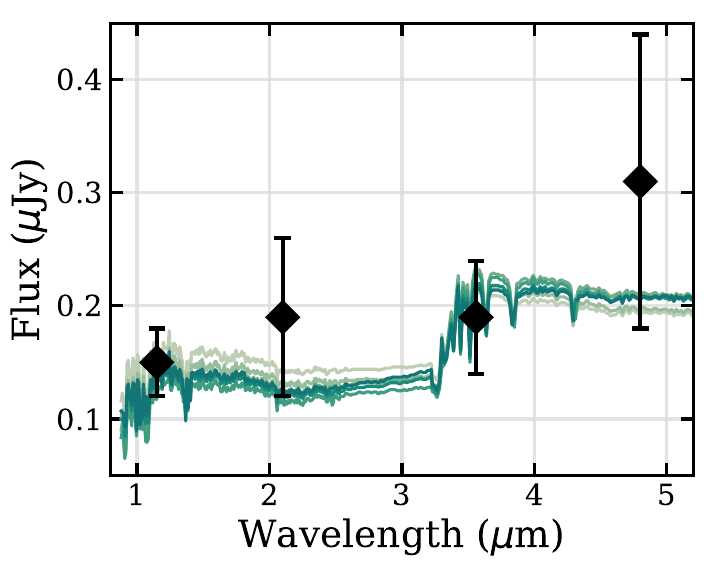}
    \caption{The flux of J1120+0641's host galaxy in the rest UV/optical compared with a set of stellar SED models to illustrate the low levels of extinction require to fit the data. The models include measurements from Table~\ref{tab:fluxes} (diamonds) compared with a set of stellar SED models (shades of green) to illustrate the low levels of extinction require to fit the data. The models include constant star formation for 350 Myr (from z $\ge$ 13), $A_V$ = 0.03; 250 Myr, 150 Myr, and 85 Myr, $A_V$ = 0.1, 0.15, and 0.25 respectively; and 350 Myr with the SFR for the most recent 10 Myr doubled and tripled, $A_V$ = 0.15 and 0.25 respectively. To simplify the figure, only the continua (no emission lines) are displayed. }
    \label{fig:fluxes}
\end{figure}

If the flux from the older stellar population (visible to JWST) is similarly obscured to the newly forming population, the mass will be underestimated; this may have contributed to the large positive uncertainty in the \texttt{Prospector} models. We can show more directly that this is unlikely to be a major issue by fitting the combined galaxy flux measurements from \citet{Stone2024} and \cite{Marshall2024} (reported in Table~\ref{tab:flux}). Figure~\ref{fig:fluxes} shows these fits: models with a broad range of star formation histories can fit the data without invoking large extinction. In addition, there is no evidence from the photometry for extinction sufficiently strong to result in a significant underestimate of the galaxy mass.

\begin{deluxetable}{ccc}
\label{tab:fluxes}
\tablecaption{Flux Measurements}
\label{tab:flux}
\tablehead{
\colhead{wavelength($\mu$m)} & 
\colhead{Flux ($\mu$Jy)} & \colhead{error} 
}
\startdata
1.15   & 0.15   & 0.03 \\
2.05  & 0.19   & 0.07 \\
3.56   & 0.19   & 0.05 \\
4.80   & 0.31   & 0.13 \\
\enddata

\end{deluxetable}

 \subsubsection{Potential total host galaxy mass}\label{sec:total}

Combining the $1-2\times10^9$ M$_{\odot}$ of stars forming in the ULIRG process (Section \ref{sec:ulirg}), the $\sim 3.2 \times 10^9$ M$_{\odot}$ of older stars (Section \ref{sec:galaxy}), and $(2.7\pm0.5)\times10^9$ M$_{\odot}$ in the merging companion discovered by \cite{Marshall2024}, the total potential stellar mass for the host galaxy becomes $7.6^{+2.9}_{-1.9}\times10^9$ M$_\odot$, assuming a local, Chabrier/Kroupa IMF for the stellar population(s).

We note that there are arguments that a top heavy IMF is more appropriate \citep{Chon2022, Steinhardt2023, Bate2025}, because the warmer cosmic background radiation during the Epoch of Reionization (e.g., $\sim$ 30 K at z = 10) warms the ISM, and if the ISM is metal-rich it radiates this heat away inefficiently. The warmer gas may inhibit the formation of low-mass stars and brown dwarfs. For a massive galaxy at $z=7$ like the J1120+0641 host, the typical metallicity is $\sim$ 0.1 Z$_\odot$ \citep{Curti2024}. The models of \cite{Bate2025} then predict a reduction of the host stellar mass of $\sim25$\% compared to a local IMF. We proceed with the mass values for a local IMF rather than a top-heavy IMF, however, as this gives the highest masses and therefore the most conservative estimate of $M_{\mathrm{BH}}/M_*$ at $z=0$.   

Another contribution to the host galaxy stellar mass is its atomic and molecular gas reservoir, which does not contribute to its luminosity as observed by JWST. Values for the mass of this gas are vary significantly. \citet{Neeleman2021} provide an estimate of $0.8^{+4.8}_{-0.3} \times 10^{10}$ M$_\odot$ by converting the far-IR flux density from ALMA to a dust mass and assuming a dust-to-gas mass ratio. They also estimate a mass of $3.8^{+3.5}_{-1.8} \times 10^{10}$ M$_\odot$ directly from the [C~II] luminosity. \citet{Venemans2017} derive upper limits for the gas mass using two mm-wave lines with ALMA: $< 3 \times 10^{10}$ from an upper limit on the CO(2-1) line flux; and $<  2 \times 10^{10}$ M$_\odot$ from an upper limit on  the CO(7-6) line flux. All of these values are consistent with $2 \times 10^{10}$ M$_\odot$, which we adopt. Only a fraction of this gas will eventually be converted into stars, governed by the star formation efficiency (SFE). \citet{Polzin2024} find SFEs of a few \%, independent of metallicity. \citet{Kim2021} find possible SFEs of a few \% to about 10\%. \citet{Andalman2025} run simulations for $z = 9$ where the high gas density can favor increased SFE, and typically find values between 10\% and 20\%. However, estimates of the SFE are highly variable in general, depending on the assumed boundary conditions and approach. To be conservative and avoid underestimating the mass, we adopt a fiducial value of 20\% for J1120+0641, in which case the gas presently in the galaxy could eventually form about $4 \times 10^9$ M$_\odot$ of stars. For the effect of other choices of SFE on the final mass, see Table \ref{tab:uncertainties}.

Assuming that the interacting galaxy is subsumed into the host, and adding the other potential stellar contributions, we project that---not yet accounting for accretion of any galaxies in the field---the existing resources could lead to a host of about $1.3 \times 10^{10}$ M$_\odot$, about nine times the mass of the black hole and  a factor of $\sim$ 100 smaller than the standard Magorrian ratio ($\sim0.001$, for a predicted host mass of $\sim1.5\times10^{12}$ M$_{\odot}$). The effects of the uncertainty on the stellar mass and any changes to the adopted gas mass or SFE are reported in Table \ref{tab:uncertainties}: there is approximately a factor of 2 error on this value.

\begin{deluxetable*}{lccccc}
\tablecaption{Uncertainties on the final ($z=0$) mass of ULAS J1120+0641's host galaxy. \label{tab:uncertainties}}
\tablehead{
\colhead{Source of uncertainty} & \colhead{Minimum} & \colhead{Adopted} & \colhead{Maximum} & \colhead{Unit} & \colhead{Reference(s)}
}
\startdata
\multicolumn{5}{c}{\bf CURRENT HOST GALAXY}  \\
{\bf Stars} & & & & \\
Stellar mass at $z=7.08$ & 0.57 & 0.76 & 1.05 & $\times 10^{10}$ M$_{\odot}$ & M24 \\ [8pt]
{\bf Gas/Future Stars} & & & & \\
Stellar mass forming in current starburst & 0.05 & 0.10 & 0.20 & $\times 10^{10}$ M$_{\odot}$ & this work \\ [2pt]
Gas mass at $z=7.08$ & 0.50 & 2.00 & 3.00 & $\times 10^{10}$ M$_{\odot}$ & N21, V17 \\ [-2pt]
$\times$ Star formation efficiency to $z=0$ & 0.10 & 0.20 & 0.40 & & An25, K21, P24 \\ [-2pt]
= Stellar mass formed by $z=0$ & 0.05 & 0.40 & 1.20 & $\times 10^{10}$ M$_{\odot}$ & \\ [8pt]
{\bf Stellar Mass at $z=0$} & {\bf 0.67} & {\bf 1.26} & {\bf 2.45} & $\times 10^{10}$ M$_{\odot}$ & \\
\hline
\multicolumn{5}{c}{\bf SATELLITE GALAXIES (QUASAR FIELD ONLY)} \\
{\bf Stars} & & & & \\
Stellar mass detected & 0.20 & 0.71 & 1.21 & $\times 10^{10}$ M$_{\odot}$ & this work \\ [2pt]
+ Stellar mass missed (incompleteness) & 0.18 & 0.35 & 0.53 & $\times 10^{10}$ M$_{\odot}$ & \\ [-2pt]
+ Stellar mass missed (no [O~III])& 0.19 & 0.52 & 0.85 & $\times 10^{10}$ M$_{\odot}$ & \\ [2pt]
= Total stellar mass ($z=7.08$) & 0.57 & 1.58 & 2.59 & $\times 10^{10}$ M$_{\odot}$ &\\  [8pt]
{\bf Gas/Future Stars} & & & & \\
$M_{\mathrm{gas}}/M_*$ at $z=7.08$ & 5 & 10 & 20 & & Al25, H22 \\ [-2pt]
$\hookrightarrow$ Gas mass at $z=7.08$ & 2.85 & 15.8 & 51.8 & $\times 10^{10}$ M$_{\odot}$ & \\ [-2pt]
$\times$ Star formation efficiency to $z=0$ & 0.1 & 0.2 & 0.4 & & An25, K21, P24\\ [-2pt]
= Stellar mass formed by $z=0$ & 0.29 & 3.15 & 20.7 & $\times 10^{10}$ M$_{\odot}$ & \\  [8pt]
{\bf Stellar Mass at $z=0$} & {\bf 0.86} & {\bf 4.73} & \bf{23.3} & $\times 10^{10}$ M$_{\odot}$ & \\
\hline
\multicolumn{5}{c}{\bf TOTAL} \\
{\bf Host Stellar Mass at $z=0$:} & {\bf 1.53} & {\bf 5.99} & {\bf 25.8} & $\times 10^{10}$ M$_{\odot}$ & \\
\enddata
\tablecomments{References: \cite{Algera2025} (Al25), \cite{Andalman2025} (An25), \cite{Heintz2022} (H22), \cite{Kim2021} (K21), \cite{Marshall2024} (M24), \cite{Neeleman2021} (N21), \cite{Polzin2024} (P24), \cite{Venemans2017} (V17).}
\end{deluxetable*}

\subsection{The Future of ULAS J1120+0641}\label{sec:future}

The host galaxy of J1120+0641 can grow via mergers with its surrounding galaxies between $z=7.08$ and $z=0$. Here, we estimate the total mass available to be accreted and to form stars.

As discussed in Section \ref{sec:sample}, the two SED fitting codes we use return very different masses for the surrounding galaxies due to our lack of wavelength coverage in the (rest-frame) red and near-IR. \texttt{Bagpipes}, which prefers to fit extremely young stellar populations (mass-weighted age $\lesssim20$ Myr) for almost all sources, returns masses between $2\times10^7$ and $5\times10^8$ M$_{\odot}$ for the high-confidence candidates. \texttt{Prospector}, which fits more evolved stellar populations with smaller emission-line contributions, yields $2\times10^8$ to $3\times10^9$ M$_{\odot}$ for the same objects, approximately an order of magnitude higher in most cases. The masses from \texttt{Prospector} at lower redshift and on relatively old galaxies have been shown to agree well on average with dynamical masses \citep{Cappellari2023}. Given comparable initial conditions, \texttt{Prospector} and \texttt{Bagpipes} return very similar mass estimates \citep{Pacifici2023}. That is, the differences in mass estimates do not reflect uncertainties in the modeling programs. Indeed, we find that if \texttt{Bagpipes} is forced to fit an older stellar population, it returns very similar results to \texttt{Prospector} (albeit with a larger $\chi$-squared for the fit). The \texttt{Prospector} masses are traceable to dynamical masses if the galaxy masses are dominated by relatively old stellar populations and the \texttt{Bagpipes} masses represent the situation if very young stars are dominant.

As such, to estimate the total mass of our candidate [O~III] emitters and its uncertainty, we adopt the \texttt{Bagpipes} value as a lower limit on the stellar mass and the \texttt{Prospector} value as an upper limit, and adopt the median value going forward. Table \ref{tab:uncertainties} reports the uncertainty that this choice, and all others described below, have on the total inferred mass. 

Under this choice, the narrow-band candidates across both fields have a total stellar mass of approximately $1.2\times10^{10}$ M$_{\odot}$ ($2.2\times10^{11}$ M$_{\odot}$ if the medium-confidence sources, two of which are fit to dubiously high masses, are included). $(7.1\pm5.0)\times10^{9}$ M$_{\odot}$ of galaxies lies in the quasar field. All galaxies in the quasar field are high-confidence sources (see Figure \ref{fig:candidates}).

It is clear from Figure \ref{fig:counts} that the adjacent field away from the quasar is not a true blank field. The similar density of galaxies observed in the quasar and adjacent fields implies that both fields trace an overdensity spanning $\gtrsim 1$ pMpc. However, even if the overdensity extends over a very large area, not all galaxies in the overdensity are likely to merge with the quasar host in the finite time between the quasar's observed redshift and the present day. \cite{Chamberlain2024} examine Illustris TNG100 galaxy pairs with similar stellar masses to our sources, and find that galaxy pairs that merge by $z=0$ generally have separations $\leq 300$ kpc, even at the highest redshift they consider ($z=6$). The candidates in the quasar field have projected distances $\lesssim 350$ kpc from the quasar, while the candidates in the adjacent field have projected distances $\gtrsim 650$ kpc from the quasar. Any galaxies off the edge of our JWST data will also have projected distances $\gtrsim 300$ kpc. It is therefore realistic and conservative to assume that only the sources in the quasar field could merge with the quasar host to increase its mass: $(7.1\pm5.0) \times10^{9}$ M$_{\odot}$ of stars.

However, some galaxies around the quasar---and therefore some mass potentially available to accrete onto the host---may be missing from our sample. The effects of incompleteness are visible at $m_{\mathrm{F150W}} \gtrsim 28$, and any galaxies without bright [O~III] will be missed by our narrow-band selection.

To assess the effects of incompleteness, we extrapolate our counts assuming the true distribution of galaxies follows the shape of the \cite{Bouwens2015} counts to the faint end (30th mag). We scale the \cite{Bouwens2015} HST counts to match our JWST counts at $m_{\mathrm{F150W}} = 26 - 28$, which yields $\log(\mathrm{N/arcmin}^2) = 0$ and $0.5$ at $m_{\mathrm{F150W}} = 28.5$ and $29.5$ respectively. This predicts that $\sim5$ galaxies with $28 < m_{\mathrm{F150W}} < 29$ and $\sim 15$ galaxies with $29 < m_{\mathrm{F150W}} < 30$ should be present in the 4.4 arcmin$^2$ quasar field. We detect 4 and 2 galaxies at these magnitude ranges, respectively, and use their masses to estimate the mass of ``missing" galaxies, yielding $3.5\pm{+1.8}\times10^{9}$ M$_{\odot}$ missed due to incompleteness: the total stellar mass in the quasar field is then $1.06\pm0.68\times10^{10}$ M$_{\odot}$.

We expect the majority of sources at $z\sim7.1$ to be highly star-forming and therefore bright in [O~III], in line with the bulk of high-redshift galaxies detected with JWST. However, it is important to understand what fraction of $z\sim7.1$ galaxies our selection technique will miss.

The JADES survey \citep{Eisenstein2023} covers a much larger area than our observations (45 arcmin$^2$) to great depth, detecting hundreds of $z\sim7$ galaxies and assigning robust photometric redshifts. JADES does not observe in F405N or F360M, two of the primary bands used in our sample selection, but does use F410M and F200W. Comparing the ratio of these two bands allows us to estimate the fraction of $z\sim7$ galaxies that lack a bright [O~III] line. In principle, any galaxy with F410M brighter than F200W will be selected by our criteria described in Section \ref{sec:candidates}. In practice, however, galaxies with brighter [O~III] are more easily fit to the quasar redshift by SED fitting codes. For this reason, the minimum observed $m_{\mathrm{F200W}} - m_{\mathrm{F410M}}$ color in our $z\sim7.1$ sample is 0.7. 

\begin{figure}
    \centering
    \includegraphics[width=8cm]{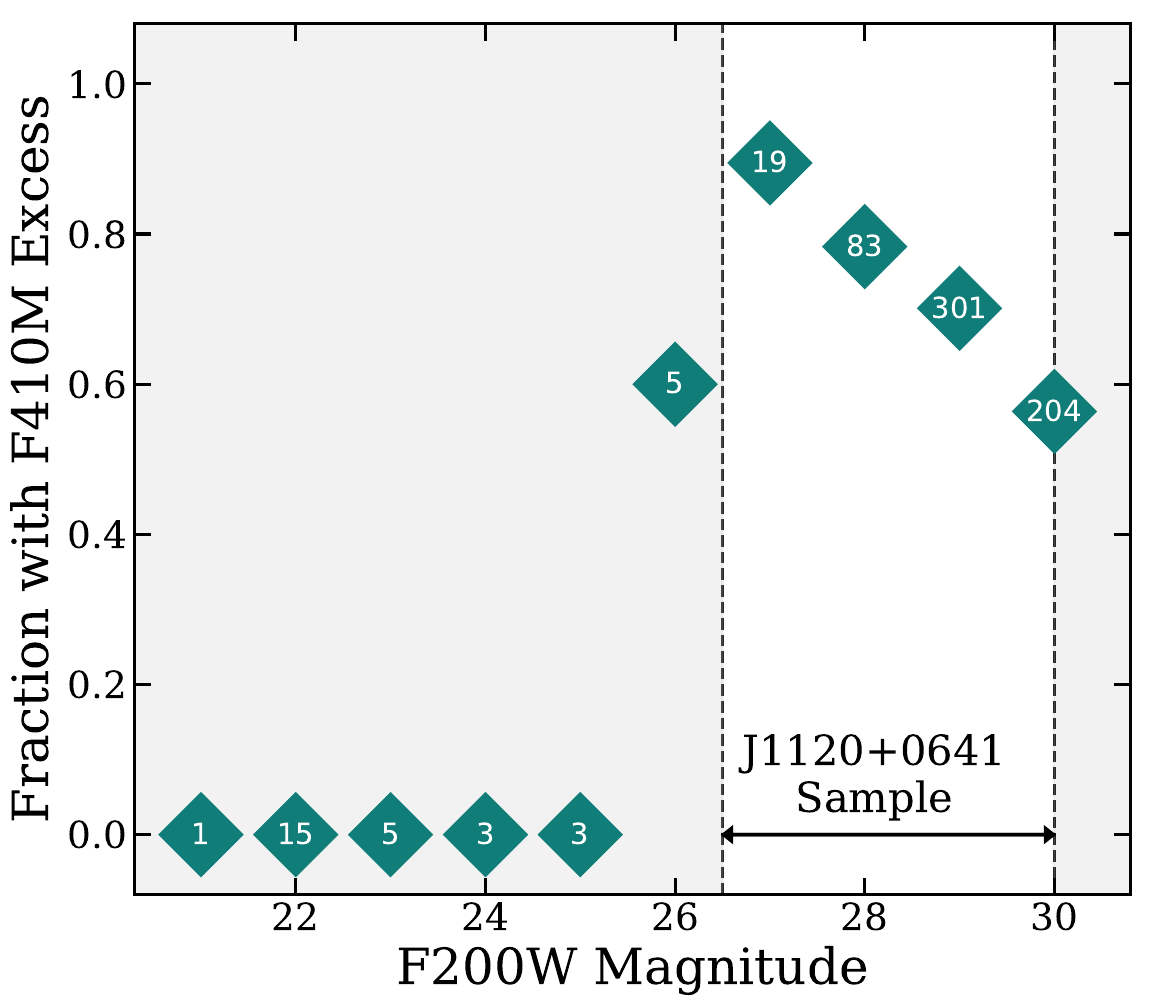}
    \caption{The fraction of $z_{\mathrm{phot}}\in[7.0, 7.5]$ galaxies in JADES with an excess in F410M relative to F200W, as a function of F200W magnitude. The number of galaxies in each bin is indicated in white text. Our brightest $z\sim7.1$ source has an F200W magnitude of 26.6: at this magnitude and fainter, almost 100\% of JADES sources have an F410M excess indicating the presence of [O~III]. Visual inspection reveals that the small number of sources at brighter magnitudes (grey shaded region), which tend not to display this excess, are almost all faint stars with poor photometric redshift fits. Essentially all galaxies at $z\sim7$ have bright lines in our bands of interest, indicating that our narrow-band selection does not leave out a significant population of line-less galaxies.}
    \label{fig:o3stats}
\end{figure}

We examine the $m_{\mathrm{F200W}} - m_{\mathrm{F410M}}$ colors of all JADES galaxies which are a) detected in both F200W and F410M, b) are fit to photometric redshifts $z \in [7.0, 7.5]$, and c) have F200W magnitudes between 26.5 and 30, to match our $z\sim7.1$ sample. 67.2\% of these JADES galaxies have $m_{\mathrm{F200W}} - m_{\mathrm{F410M}}$ colors consistent with our sample: this indicates that approximately 32.8\% of galaxies do not have bright enough [O~III] in the bands to be effectively fit by the SED fitting codes. We note that the very bright sources in JADES ($m_{\mathrm{F200W}} \lesssim 25$), almost none of which have a strong F410M excess (see Figure \ref{fig:o3stats}, are mostly misclassified stars. The population of galaxies without bright [O~III] may not follow the same spatial distribution or mass distribution as the [O~III] emitters: \cite{Champagne2025b} examined the environment of the $z=6.61$ quasar J0305-3150 and found a correlation between [O~III] equivalent width and distance from the central quasar. They also found that the Lyman-break galaxies near the quasar, without [O~III] emission, were universally older, less star-forming, and more massive. However, without more knowledge of the population of galaxies around J1120+0641 without bright [O~III], we cannot robustly correct for any difference in the populations. We therefore assume that galaxies without strong [O~III] follow the same mass distribution as the detected sources, and correct the detected stellar masses up assuming 32.8\% of the mass is missed. This adds another $5.2\pm3.3\times10^9$ M$_{\odot}$, and brings the total stellar mass in the quasar field to $1.58\pm1.01\times10^{10}$ M$_{\odot}$. 

With a robust estimate of the total stellar mass in the satellite galaxies, we finally need to account for the gas present in the satellite galaxies, some fraction of which will form stars. While no comprehensive surveys from e.g. ALMA exist at the high redshift and very low mass of our sample galaxies, galaxies of slightly higher mass at the same redshift tend to have gas masses approximately an order of magnitude larger than their stellar masses \citep{Algera2025, DeGraaff2024, Heintz2022}. Assuming a gas-to-stars ratio of 10 and an SFE of 20\% (see references in Section \ref{sec:total}), the total stellar mass of $\sim1.6\times10^{10}$ M$_{\odot}$ in the surrounding galaxies implies that an additional $\sim3\times10^{10}$ M$_{\odot}$ of stars can form by $z=0$. For the effect of changing the adopted gas-to-stars ratio and SFE on the final mass, see \ref{tab:uncertainties}. Between the detected galaxies, the galaxies missed due to incompleteness, and their gas, we therefore conclude that approximately $4.8\times10^{10}$ M$_{\odot}$ of stars could possibly be added to the host galaxy's mass between its observed redshift and $z=0$ through mergers. This value is highly uncertain, dependent on the assumed gas-to-stars ratio and star formation efficiency: this mass from ``future stars" dominates the uncertainty on our final host galaxy mass (see Table \ref{tab:uncertainties}). To see the effect that different assumptions have on the derived total mass, see Table \ref{tab:uncertainties}.

\begin{figure}
    \centering
    \includegraphics[width=8cm]{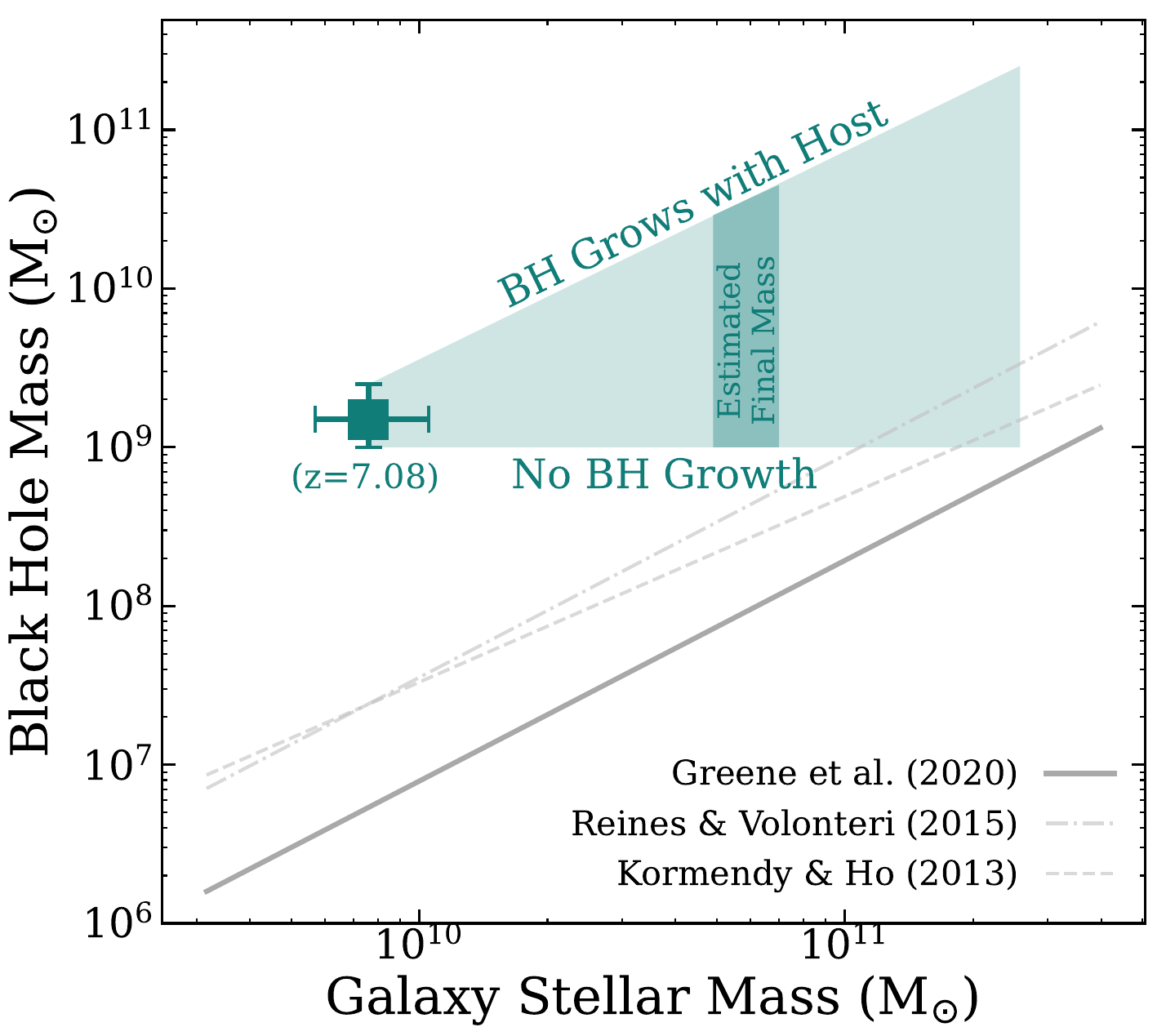}
    \caption{The future of J1120+0641's host galaxy. J1120+0641's black hole and stellar mass at $z=7.08$ is marked with a teal square. The lighter teal polygon marks the region where the host could land at $z=0$, up to the maximum possible galaxy mass we derive, $2.5\times10^{11}$ M$_{\odot}$ (see Table \ref{tab:uncertainties}). The lower boundary assumes no black hole growth, while the upper boundary assumes the black hole grows at the same rate as the host galaxy. The solid thick grey line is the local relationship between black hole and stellar mass from \cite{Greene2020}, and we show similar relations from \cite{Kormendy2013} and \cite{Reines2015} as lighter, dashed and dot-dashed lines respectively. We expect the host galaxy to grow to approximately $6\times10^{10}$ M$_{\odot}$ in stellar mass: this area of the polygon is shaded darker and labeled. With no further black hole growth and a mass of $6\times10^{10}$ M$_{\odot}$, J1120+0641 will reach an $M_{\mathrm{BH}}$-$M_*$ ratio of $\sim2.5$\% at $z=0$, still significantly above the local \cite{Greene2020} relation.}
    \label{fig:magorrian}
\end{figure}

Combining the mass of the surrounding galaxies with that of the host gives a total mass of $\sim 6.0\times10^{10}$ M$_{\odot}$. Assuming that no further black hole growth occurs, then the host of J1120+0641 will grow to $M_{\mathrm{BH}}/M_* \sim 0.025$, or 2.5\%, at $z=0$: still about a factor of 25 greater than the typical value for low-redshift galaxies of similar stellar mass \citep[$M_{\mathrm{BH}}/M_* \sim 0.001$,][]{Greene2020}. Only as we approach the upper end of our uncertainty range---still assuming no further black hole growth, and adopting the large \texttt{Prospector} masses for the satellites as well as large gas fractions and very efficient star formation---do we achieve a final stellar mass of $2.6\times10^{11}$ M$_{\odot}$, for $M_{\mathrm{BH}}/M_* \sim 0.005$ (still on the lower end of the expected stellar mass given the mass of J1120+0641's black hole). The range in $M_{\mathrm{BH}}/M_*$ that J1120+0641 could reach is shown in Figure \ref{fig:magorrian}.

Galaxies with $M_{\mathrm{BH}}/M_* \sim 2.5\%$ are essentially unheard of in the local Universe. Of the sources in \cite{Greene2020} with similar black hole masses to J1120+0641, none have stellar masses below $10^{11}$ M$_{\odot}$. The few existing examples of such extreme systems are unique cases: NGC 1277 was originally thought to have a black hole mass of $1.7\times10^{10}$ M$_{\odot}$ and $M_{\mathrm{BH}}/M_*$ of 6.3\% \citep{VanDenBosch2012}, but later modeling \citep{Emsellem2013, Graham2016} found that the dynamics of the bulge could also be explained by a bar viewed end-on, and revised the black hole mass down, closer to $1.2\times10^9$ solar masses ($M_{\mathrm{BH}}/M_* = 0.45\%$). NGC 4342 has a black hole mass of $3\times10^8$ M$_{\odot}$ \citep[$M_{\mathrm{BH}}/M_{\mathrm{bulge}} = 6.9\%$,][]{Cretton1999}, but there is evidence that it is being tidally stripped by its neighbor, NGC 4365 \citep{Blom2014}, reducing its stellar mass. M32 is another classic example of this phenomenon, tidally stripped by interactions with M31. There appears to exist no example of a $z\sim0$ galaxy with $M_{\mathrm{BH}}/M_*$ similar to that implied for J1120+0641 by the total mass around it.

It is therefore clear that the galaxies visible in the immediate surroundings of J1120+0641 are likely not sufficient to transport it onto the local $M_{\mathrm{BH}}/M_*$ relation. Sub-mm data to constrain the gas mass of these satellites, or typical low-mass galaxies at $z\sim7$, could reduce our uncertainty significantly. If the host is to grow further, it must either accrete significantly more gas from the surrounding intergalactic medium or undergo a merger of its dark matter halo with another halo not visible in our observed field. Predicting this is, of course, impossible. However, if there are many other examples where this process is invoked, it would imply that the Magorrian relation is achieved in those cases stochastically, and not in a repeatable fashion.

It's important to note that the presence of even a single massive NIRCam-dark galaxy in the vicinity of J1120+0641, similar to those discovered around the quasars J0305-3150 and J1526-2050 by \cite{FSun2025} at $z\sim 6.6$, could significantly increase the mass available to accrete onto the quasar host. No [C~II] imaging of J1120+0641 currently exists over a large enough area to confirm or refute this possibility.

It may be that the fate of J1120+0641 is in fact to evolve into a SMBH with very little host galaxy. Then the absence of known similar cases in the local Universe may be due to selection effects \citep[but note the example of such a system at intermediate redshift presented by][]{Schramm2019}. At the end of the current quasar phase, the SMBH would likely not be elevated to quasar luminosity again, due to the apparent dearth of gas around the galaxy to accrete, and therefore evolve quietly to low redshift. A quiescent SMBH in a low luminosity galaxy might escape detection altogether if it were sufficiently distant. 

\subsection{The Past of ULAS J1120+0641}

It is emerging from JWST observations that, at high redshift, there are many examples of  black holes overmassive relative to their host galaxies \citep[e.g.,][]{Harikane2023, Pacucci2023, Ubler2023, Kokorev2023, Goulding2023, Furtak2024,  Juod2024, Natarajan2024}. For the most massive cases (log($M_{\mathrm{BH}}/M_\odot) > 8$), the available sample sizes are inadequate to  show whether this behavior is typical or is confined to a subsample that is being found preferentially \citep{YSun2025}. However, there  are individual cases that stand out; among these, J1120+0641, with $M_{\mathrm{BH}}/M_{*} \sim 0.3$,  is unique in the amount and quality of data to evaluate this  behavior. As a result, concerns about the validity of the mass measurements of the black hole \citep{Lupi2024a, King2024} (and of the host galaxy) are greatly reduced and we can confidently explore the implications for its past evolution.

The large black hole mass of J1120+0641 only 750 Gyr after the Big Bang indicates a massive seed black hole. To reach this mass via constant Eddington accretion requires a seed mass of $1.26\times10^4$ M$_{\odot}$ if the accretion begins at $z=13$ and 4400 M$_{\odot}$ if the accretion begins at $z=20$. These seed masses resemble those predicted by direct-collapse theory, where a pristine gas cloud is prevented from fragmenting by Lyman-Werner radiation from nearby stars \citep[e.g.,][]{Agarwal2013, Chiaki2023}, though a number of physical challenges to this formation pathway exist \citep{Bhowmick2022a, Bhowmick2022b, Dunn2018, Smith2017}. 

If this formation pathway is possible, J1120+0641 may be an example of the evolution of an Overly Massive Black hole Galaxy \citep[OMBG,][]{Agarwal2013, Natarajan2017, Scoggins2023, Scoggins2024}, where a massive black hole forms via direct collapse in a satellite subhalo of a star-forming halo which provides the Lyman-Werner radiation flux. These systems are predicted to have very large $M_{\mathrm{BH}}/M_*$ (greater than 1) until their host halo merges with the nearby larger halo, at which point the $M_{\mathrm{BH}}/M_*$ ratio is sub-unity, but still much greater than observed in the local Universe. 

Alternatively, with episodes of  super-Eddington accretion a large  black hole mass can be achieved with a lower-mass seed and with a different accretion history \citep[see e.g.][]{Hu2022, Lupi2024b, Trinca2024}. In the detailed simulation by \citet{Lupi2024b}, the super-Eddington phase is delayed since supernova feedback agitates the interstellar gas and prevents inflow and accretion. However, once the galaxy grows in mass to $\sim$ 10$^{10}$ M$_\odot$, accretion can occur very rapidly and grow the black hole by orders of magnitude. A similar result is obtained in the independent simulation by \citet{Husko2025}. This hypothesis would make cases like  ULAS J1120  rare, particularly at higher redshifts before galaxies have grown to sufficient mass. In comparison, \citet{Trinca2024} propose that super-Eddington acretion is triggered by large scale dynamical events such as major mergers that can occur over an extended period of time. In this type of model there are bursty episodes of super-Eddington accretion throughout the early life of the host galaxy.

\section{Summary and Conclusions} \label{sec:summary}

Using new, deep JWST/NIRCam imaging, we searched for [O~III] emitters around the $z=7.08$ quasar ULAS J1120+0641 to constrain the density of the quasar's environment. We use a narrow-band selection technique combined with modeling of the broadband UV/optical SED to obtain robust redshifts and masses.

\begin{itemize}
    \item We place 19 galaxies at the quasar redshift with high confidence (3 at medium confidence), 13 of which lie in the quasar field, within a projected distance of 350 kpc. The remaining 6 high-confidence candidates and the 3 medium-confidence candidates lie in the adjacent field, with projected distances from the quasar $>650$ kpc. 
    \item We determine that two of the Lyman-break galaxies detected in the quasar field by \cite{Simpson2014} are not directly associated with the quasar. LBG1 lies at $z=6.95$ and LBG2 lies at $z=2.01$.
    \item J1120+0641 lies in an overdensity of galaxies spanning more than 1 proper Mpc. Both the quasar field and adjacent field are significantly overdense in $z \in [7.05, 7.15]$ sources down to our completeness limit, most of which could not be identified with HST alone.
    \item We combine multiple observations of the quasar to better constrain its properties, arriving at a current black hole mass $M_{\mathrm{BH}} \sim 1.5\times10^9$ M$_{\odot}$. Masses derived from single-epoch spectroscopy, modeling of the accretion disk, and Eddington ratio arguments are in rough agreement. This high mass less than 1 Gyr after the Big Bang likely requires  a massive black hole seed and/or significant episodes of super-Eddington accretion 
    \item We constrain the properties of the host galaxy by combining data from \cite{Marshall2024}, \cite{Yue2024}, \cite{Stone2024}, and \cite{Bosman2024}. The host resembles a ULIRG, with high levels of dust-obscured star formation in the galaxy's center, likely ignited by an ongoing merger. The two merging components of the quasar host have a total stellar mass of $\sim7.6\times10^{9}$ M$_{\odot}$, with an ongoing starburst, and gas supply with the potential to increase this mass.
    \item After accounting for incompleteness and selection effects, the stellar mass in the galaxies within $\sim350$ kpc (projected) of the quasar is $\sim(1.6\pm1.0)\times10^{10}$ M$_{\odot}$. If all the galaxies in the overdensity within $\sim350$ kpc of the quasar merge with the host galaxy by $z=0$ and form additional stars, the host of J1120+0641 will reach a final stellar mass of $\sim6\times10^{10}$ M$_{\odot}$ (with a large uncertainty dominated by the error in the gas masses). If the black hole grows no further, this corresponds to a black hole mass--stellar mass ratio of $\sim2.5$\%: significantly larger than any known galaxy in the local Universe that has not lost significant stellar mass to tidal stripping. Any further growth of the host galaxy will be stochastic, through e.g. unexpectedly high accretion of gas from the intergalactic medium and/or a merger with another halo.
\end{itemize}

\begin{acknowledgments}

Work on this paper was supported in part by grant 80NSSC18K0555, from NASA Goddard Space Flight Center to the University of Arizona. MS acknowledges support from the National Science Foundation Graduate Research Fellowship under Grant No. DGE-2137419, and thanks Jake Helton for his assistance with \texttt{Bagpipes} and Maria Pudoka for helpful comments. The JWST data presented in this paper were obtained from the Mikulski Archive for Space Telescopes (MAST) at the Space Telescope Science Institute. The observations can be accessed via\dataset[DOI.]{http://dx.doi.org/10.17909/vvcf-1d88}

\end{acknowledgments}

\vspace{5mm}
\facilities{JWST(NIRCam, NIRSpec)}

\software{Astropy \citep{Astropy2013,Astropy2018}, Bagpipes                \citep{Carnall2018}, Matplotlib \citep{Hunter2007},              NumPy \citep{VanDerWalt2011}, photutils                          \citep{Bradley2022}, Prospector \citep{Johnson2021}
         }

\appendix

\restartappendixnumbering

\section{Galaxy images}

\begin{figure*}
    \centering
    \includegraphics[width=\textwidth]{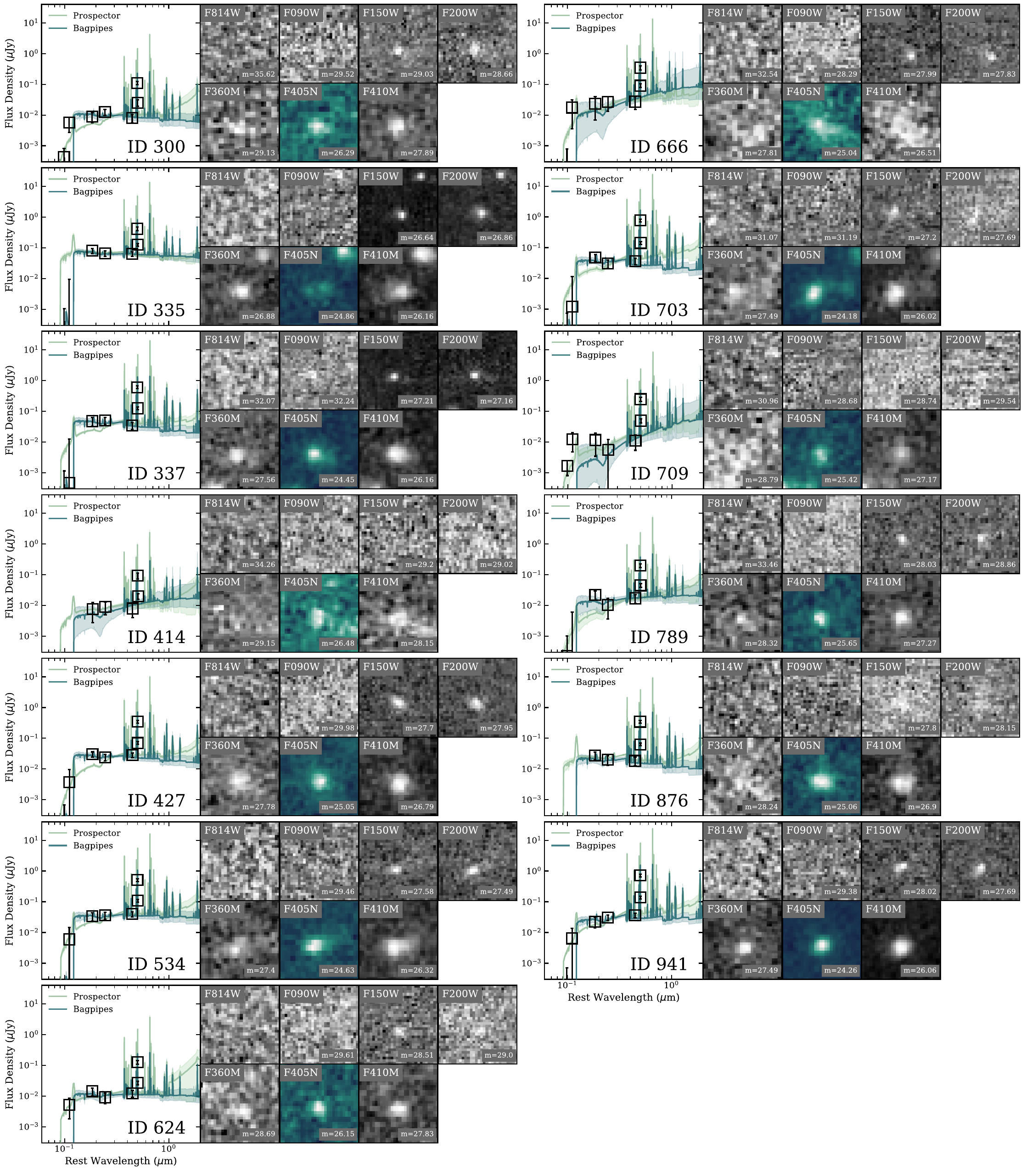}
    \caption{The thirteen $z\sim7.1$ sources detected in the field around J1120+0641 (in order of ID, see Table \ref{tab:sample}). Each source's spectral energy distribution is shown in the rest-frame in the left panel, along with the \texttt{Prospector} (light green) and \texttt{Bagpipes} (dark blue) best-fit spectra and 16-84\% confidence intervals. At right are the source's thumbnails in F814W and our JWST filter set. The narrow-band filter F405N is shown in teal. Each band with a positive measured flux has the magnitude in the band listed in the lower right corner of the image.}
    \label{fig:appendix1}
\end{figure*}

\begin{figure*}
    \centering
    \includegraphics[width=15cm]{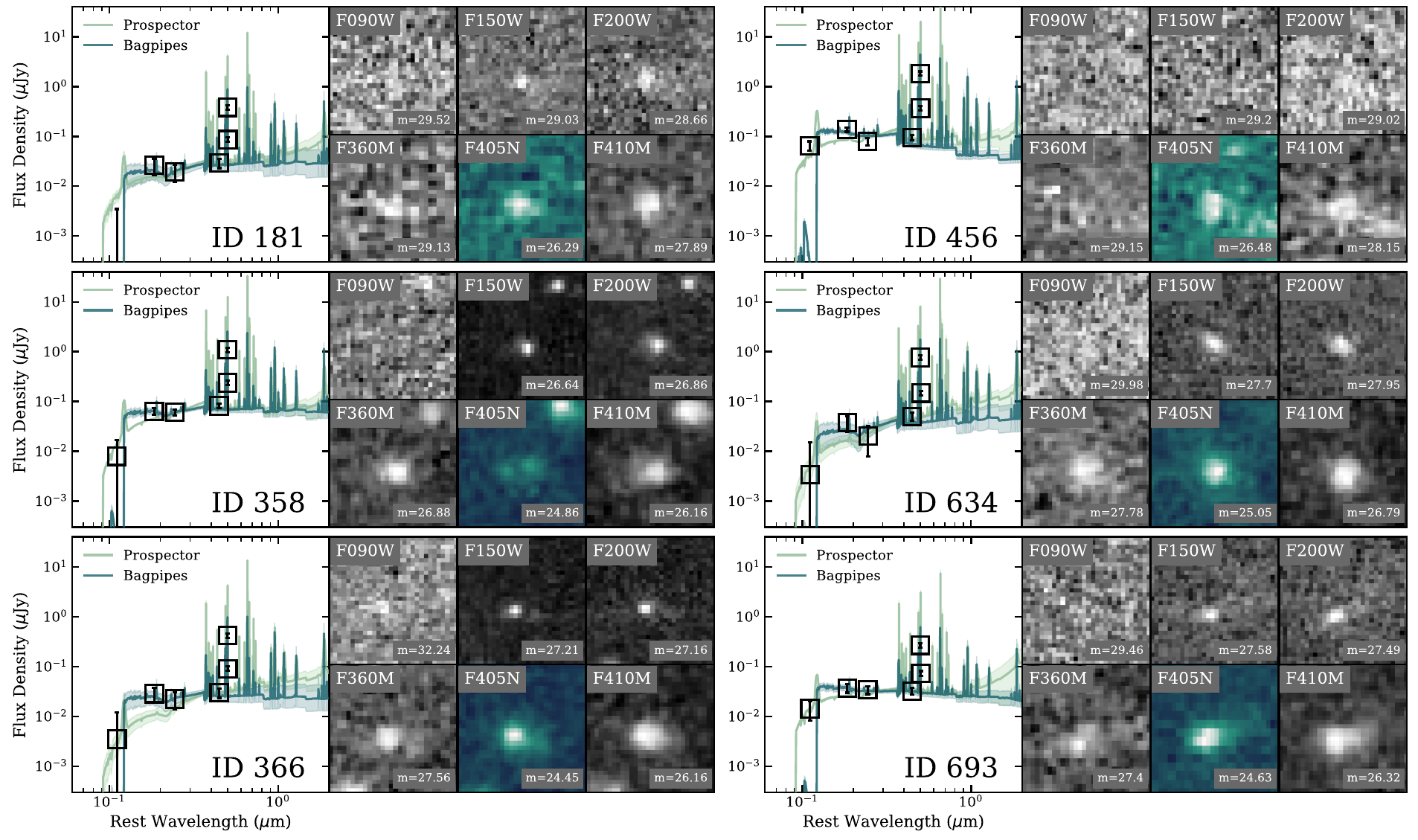}
    \caption{The six high-confidence $z\sim7.1$ sources detected in the adjacent field, as in Figure \ref{fig:appendix1}. Because this field is offset from the quasar, these sources do not have archival F814W data.}
    \label{fig:appendix2}
\end{figure*}

\begin{figure*}
    \centering
    \includegraphics[width=7.5cm]{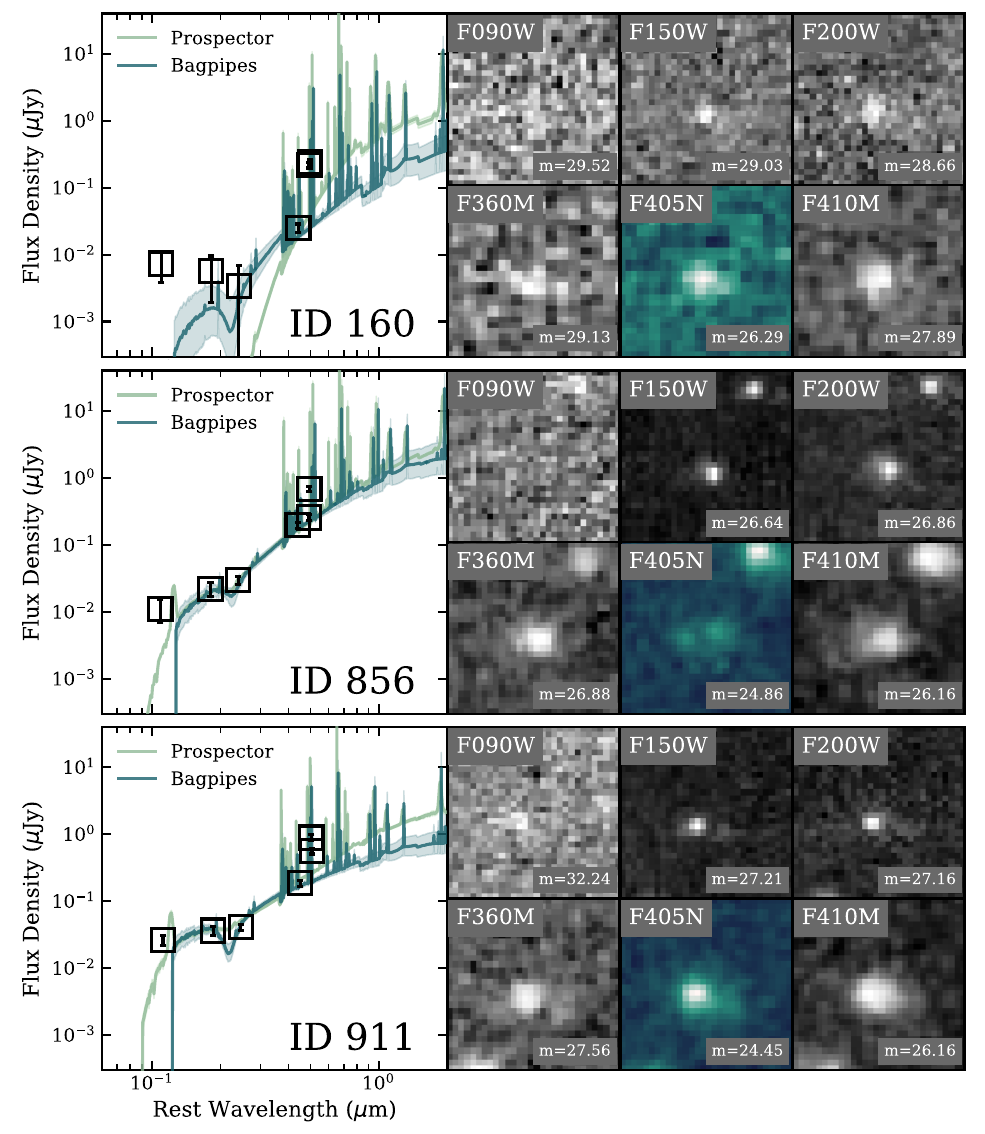}
    \caption{The three medium-confidence $z\sim7.1$ sources detected in the adjacent field, as in Figure \ref{fig:appendix1}. Unlike the high-confidence sources in Figures \ref{fig:appendix1} and \ref{fig:appendix2}, these sources are anomalously bright in the redder bands, leading to steep spectral slopes and high best-fit masses (see Table \ref{tab:sample}).}
    \label{fig:appendix3}
\end{figure*}

\eject

\section{Quasar luminosity}

\begin{figure}
    \centering
    \includegraphics[width=9cm]{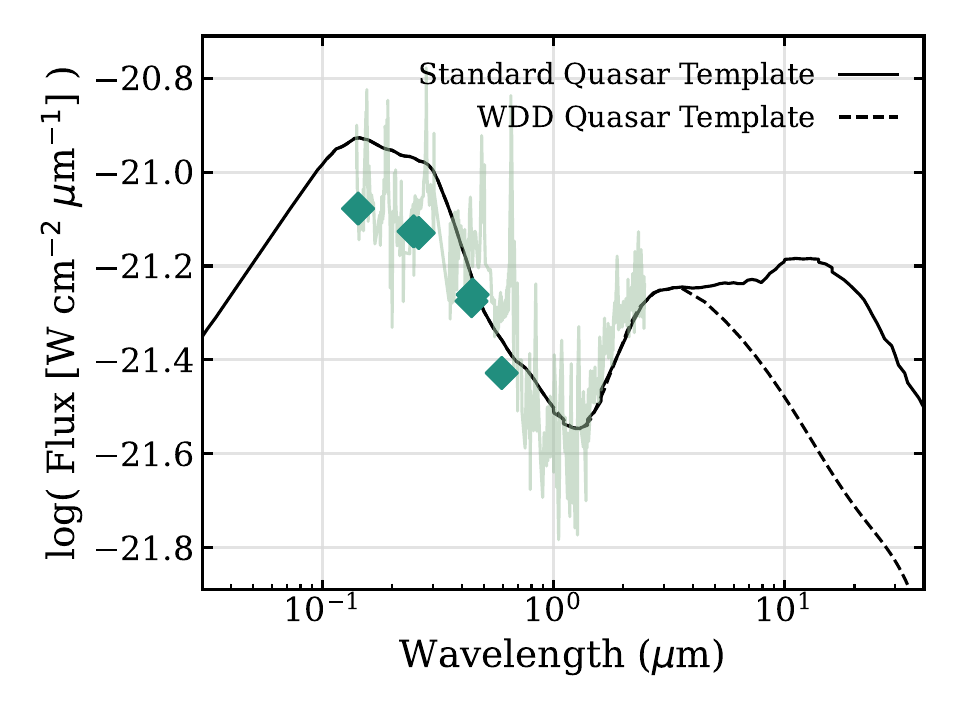}
    \caption{The SED of J1120+0641, shifted to the rest frame. The solid black line is the template SED for a ``normal'' quasar, while the dashed line shows the behavior of a warm dust deficient source \citep[which is identical at $\lambda < 3$ $\mu$m][]{Lyu2017}. A composite spectrum of J1120+0641 from \cite{Mortlock2011}, \cite{Marshall2024}, and \cite{Bosman2024} is shown in light green. The green diamonds are photometry of the quasar from \cite{Stone2024} and \cite{Yue2024}. The template has been fit to the spectrum by least squares in log space.}
  \label{fig:sed}
\end{figure}

Our determination of the luminosity of J1120+0641 is centered on Figure~\ref{fig:sed}. The $\lambda F_\lambda$ presentation shows relative luminosity for different wavelengths directly. The luminosity largely emerges between 0.05 and 30 $\mu$m. The spectra and photometry cover a significant fraction of this wavelength range, except for $\lambda > 3 \mu$m. The extinction to the BLR is minimal \citep{Bosman2024}, so the SED short of 0.1 $\mu$m should be reasonably accurate.  We therefore estimate the luminosity for the normal and WDD cases \citet{Lyu2017} by integrating the output for the appropriate SED, finding $7.4 \times 10^{13}$ L$_\odot$ and $6.6\times 10^{13}$ L$_\odot$, respectively for ``normal'' and WDD. Extrapolating from Figure 12 in \citet{Lyu2017}, it appears that there is about a 30\% probability that the WDD template is appropriate. That is, the most likely luminosity is $\sim$ $7 \times 10^{13}$ L$_\odot$, or $2.7 \times 10^{47}$ ergs s$^{-1}$. . 

\bibliography{bibliography}{}
\bibliographystyle{aasjournal}

\end{document}